\RequirePackage{rotating}

\documentclass[twocolumn]{aastex631}
\usepackage{graphicx}
\usepackage{epsfig}
\usepackage{natbib}
\usepackage{multirow}
\usepackage{amsmath}
\usepackage{textcomp}
\usepackage{rotating}

\usepackage{graphicx}
\usepackage{comment}

\newcommand{\km}{{\rm\thinspace km}}

\newcommand{\s}{{\rm\thinspace s}}

\newcommand{\kmps}{\hbox{$\km\s^{-1}\,$}}

\newcommand{\lya}{Ly$\alpha$}

\newcommand{\hi}{H\thinspace{\sc i}}

\newcommand{\oi}{O\thinspace{\sc i}}

\newcommand{\sitwo}{Si\thinspace{\sc ii}}
\newcommand{\sithree}{Si\thinspace{\sc iii}}
\newcommand{\sifour}{Si\thinspace{\sc iv}}
\newcommand{\stwo}{S\thinspace{\sc ii}}

\newcommand{\sfour}{S\thinspace{\sc iv}}
\newcommand{\sifive}{Si\thinspace{\sc v}}
\newcommand{\siiv}{Si\thinspace{\sc iv}}
\newcommand{\ci}{C\thinspace{\sc ii}}
\newcommand{\cii}{C\thinspace{\sc ii}}

\newcommand{\nv}{N\thinspace{\sc v}}
\newcommand{\nwon}{N\thinspace{\sc i}}

\usepackage{tablefootnote}

\begin{document}

\title{CLASSY X: Highlighting Differences Between Partial Covering and Semi-Analytic Modeling in the Estimate of Galactic Outflow Properties}

\author[0009-0002-9932-4461]{Mason Huberty}
\affiliation{Minnesota Institute for Astrophysics, University of Minnesota, 116 Church Street SE, Minneapolis, MN 55455, USA}
\author[0000-0003-4166-2855]{Cody Carr}
\affiliation{Center for Cosmology and Computational Astrophysics, Institute for Advanced Study in Physics, Zhejiang University, Hangzhou 310058,  China}
\affiliation{Institute of Astronomy, School of Physics, Zhejiang University, Hangzhou 310058,  China}
\author[0000-0002-9136-8876]{Claudia Scarlata}

\affiliation{Minnesota Institute for Astrophysics, University of Minnesota, 116 Church Street SE, Minneapolis, MN 55455, USA}
\author[0000-0003-1127-7497]{Timothy Heckman}

\affiliation{Center for Astrophysical Sciences, Department of Physics \& Astronomy, Johns Hopkins University, Baltimore, MD 21218, USA}
\author[0000-0002-6586-4446]{Alaina Henry}

\affiliation{Space Telescope Science Institute, 3700 San Martin Drive, Baltimore, MD 21218, USA}

\affiliation{Center for Astrophysical Sciences, Department of Physics \& Astronomy, Johns Hopkins University, Baltimore, MD 21218, USA}
\author[0000-0002-9217-7051]{Xinfeng Xu}
\affiliation{Department of Physics and Astronomy, Northwestern University, 2145 Sheridan Road, Evanston, IL, 60208, USA.
}
\affiliation{Center for Interdisciplinary Exploration and Research in Astrophysics (CIERA), Northwestern University, 1800 Sherman Avenue, Evanston, IL, 60201, USA.
}
\author[0000-0002-2644-3518]{Karla Z. Arellano-C\'{o}rdova}
\affiliation{Institute for Astronomy, University of Edinburgh, Royal Observatory, Edinburgh, EH9 3HJ, UK}
\affiliation{Department of Astronomy, The University of Texas at Austin, 2515 Speedway, Stop C1400, Austin, TX 78712, USA}

\author[0000-0002-4153-053X]{Danielle A. Berg}
\affiliation{Department of Astronomy, The University of Texas at Austin, 2515 Speedway, Stop C1400, Austin, TX 78712, USA}

\author[0000-0003-3458-2275]{St\'{e}phane Charlot}
\affiliation{Sorbonne Universit\'{e}, CNRS, UMR7095, Institut d'Astrophysique de Paris, F-75014, Paris, France}
\author[0000-0002-0302-2577]{John Chisholm}

\affiliation{Department of Astronomy, The University of Texas at Austin, 2515 Speedway, Stop C1400, Austin, TX 78712, USA}
\author[0000-0002-5659-4974]{Simon Gazagnes}
\affiliation{Department of Astronomy, The University of Texas at Austin, 2515 Speedway, Stop C1400, Austin, TX 78712, USA}
\author[0000-0001-8587-218X]{Matthew Hayes}

\affiliation{Stockholm University, Department of Astronomy and Oskar Klein Centre for Cosmoparticle Physics, AlbaNova University Centre, SE-10691, Stockholm, Sweden}

\author[0000-0003-3424-3230]{Weida Hu}
\affiliation{Department of Physics and Astronomy, Texas A\&M University, College Station, TX 77843-4242, USA; George P. and Cynthia Woods Mitchell Institute for Fundamental Physics and Astronomy, Texas A\&M University, College Station, TX 77843-4242, USA}
\author[0000-0003-4372-2006]{Bethan L. James}

\affiliation{AURA for ESA, Space Telescope Science Institute, 3700 San Martin Drive, Baltimore, MD 21218, USA}

\author[0000-0002-3959-6572]{R. Michael Jennings}

\affiliation{Center for Astrophysical Sciences, Department of Physics \& Astronomy, Johns Hopkins University, Baltimore, MD 21218, USA}

\author[0000-0003-2685-4488]{Claus Leitherer}

\affiliation{Space Telescope Science Institute, 3700 San Martin Drive, Baltimore, MD 21218, USA}

\author[0000-0001-9189-7818]{Crystal L. Martin}
\affiliation{Department of Physics, University of California, Santa Barbara, Santa Barbara, CA 93106, USA}

\author[0000-0003-2589-762X]{Matilde Mingozzi}

\affiliation{Space Telescope Science Institute, 3700 San Martin Drive, Baltimore, MD 21218, USA}
\author[0000-0003-0605-8732]{Evan D. Skillman}

\affiliation{Minnesota Institute for Astrophysics, University of Minnesota, 116 Church Street SE, Minneapolis, MN 55455, USA}
\author[0000-0001-6958-7856]{Yuma Sugahara}

\affiliation{Institute for Cosmic Ray Research, The University of Tokyo, Kashiwa-no-ha, Kashiwa 277-8582, Japan}

\affiliation{National Astronomical Observatory of Japan, 2-21-1 Osawa, Mitaka, Tokyo 181-8588, Japan}

\affiliation{Waseda Research Institute for Science and Engineering, Faculty of Science and Engineering, Waseda University, 3-4-1, Okubo, Shinjuku, Tokyo 169-8555, Japan}

\begin{abstract}
Feedback driven massive outflows play a crucial role in galaxy evolution by regulating star formation and influencing the dynamics of surrounding media.  
Extracting outflow properties from spectral lines is a notoriously difficult process for a number of reasons, including the possibility that a substantial fraction of the outflow is carried by dense gas in a very narrow range in velocity. This gas can hide in spectra with insufficient resolution. Empirically motivated analysis based on the Apparent Optical Depth method, commonly used in the literature, neglects the contribution of this gas, and may therefore underestimate the true gas column density.  More complex semi-analytical line transfer (e.g., SALT) models, on the other hand, allow for the presence of this gas by modeling the radial density and velocity of the outflows as power laws. Here we compare the two approaches to quantify the uncertainties in the inferences of outflow properties based on 1-D “down-the-barrel” spectra, using the UV spectra of the CLASSY galaxy sample. We find that empirical modeling may significantly underestimate the column densities relative to SALT analysis, particularly in the optically thick regime.  We use simulations to show that the main reason for this discrepancy is the presence of large amount of dense material at low velocities, which can be hidden by the finite spectral resolution of the data. The SALT models in turn could over-estimate the column densities if the assumed power laws of the density profiles strong are not a property of actual outflows.

\end{abstract}

\section{Introduction}\label{sec:intro}
The mass ejection of gas into the circumgalactic medium (CGM) via galactic outflows is omnipresent in star-forming galaxies \citep[e.g.,][]{shapley2003,martin2005}.
This phenomenon heavily regulates star-formation in galaxies, and thus is a large influence in the study of galactic evolution \citep[e.g.,][]{hopkins2014}.
One of the defining characteristics of outflows is the mass outflow rate, the rate at which gas (predominantly hydrogen) is removed from the star forming regions of a galaxy outwards into the CGM. Numerous techniques have been used in the past to attempt to estimate this physical parameter \citep[e.g.,][]{heckman2015,chisholm2017,xu2022}.
Mass outflow rates vary substantially across different galaxies and models, not only because of the complex geometry of the outflows that can vary from biconical to spherical \citep[e.g.,][]{carr2021}, but also because the mass outflow rate is expected to vary with radius \citep{muratov2015,chisholm2017}.

In order to estimate these outflow rates in real galaxies, we require some basic information, such as its column density, its velocity, and its distribution, that can be measured from  galaxies' spectra. ``Down-the-barrel"  ultraviolet spectroscopy (with the galaxy itself serving as the illuminating source) is one of the most widely used tools for constraining the necessary information, thanks to the wealth of emission and absorption lines of metals and hydrogen that can be identified in this wavelength  range \citep[e.g.,][]{tremonti2007,steidel2010,zhu2015,barger2016,lan2019,berg2022,li2024}. However every approach to understanding these lines comes with its own sets of assumptions and uncertainties \citep[e.g.,][]{rupke2005,weiner2009,steidel2010,rubin2010,erb2012,kornei2013,jones2013,nelson2019}.
Originating from the outflowing gas, the shape of the absorption and emission of these UV lines depends on the outflow density distribution, its geometry, dust content, and velocity field, which in turn depend on a large number of free parameters. 
Setting constraints on this complex parameter space demands high spectral resolution/high signal-to-noise data as well as modeling that is able to capture these features in a computationally viable way \citep[e.g.,][]{carr2023}. 
With its complete spectral coverage, high spectral resolution, and high S/N,  the COS Legacy Archive Spectroscopy SurveY \citep[CLASSY as presented in][]{berg2022} is one of the best data sets that provide the opportunity to use radiation transport modeling to uncover the physics of galaxy outflows.

Galaxy outflows are routinely characterized following the early work of \citet[][]{savage1991}, i.e., the so called Apparent Optical Depth method [AOD], generalized to allow a partial covering fraction as a function of projected velocity (partial covering fraction method, PCM, hereafter). \citet{xu2022} performed this analysis on the outflows identified in the CLASSY galaxies, measuring outflow velocities, gas column densities, and mass outflow rates for the sample. Having been introduced for the study of absorption systems along the lines of sight of QSOs, the AOD however, relies on a number of simplifications, including the implicit assumption of a point-like source of continuum. Additionally, these models neglect the density profile of  outflows, possibly missing substantial amount of dense material if concentrated in a small velocity range as a consequence of the finite spectral resolution of the data. Inference based on hydrodynamical simulation and full radiative transport of continuum photons, which explores the range of important physical processes, is still  time prohibitive. Semi-analytical models, such as those of \citet{scarlata2015} relax some of the assumptions of the AOD and PCM analysis, while still providing computationally feasible approaches.

In this paper we compare the PCM and SALT approaches to the modeling of the CLASSY UV spectra with the aim of quantifying the uncertainties in the inferences of outflow properties based on down-the-barrel spectroscopy. In a companion paper, we present the scaling relations derived using the more general approach introduced here.
This paper is structured as follows: In Section~\ref{sec:data}, we summarize the CLASSY data used in this paper. In Section~\ref{sec:modeling}, we review the SALT model. In Section~\ref{sec:quantity_definitions}, we discuss the fitting procedure and the derived quantities from SALT. Comparisons to the PCM modeling techniques are included in Section~\ref{sec:comparison}. 
Our conclusions are outlined in Section~\ref{sec:discussion}. 

This paper assumes a cosmology of $H_0=70$ km s$^{-1}$ Mpc$^{-1}, \Omega_m=0.3,$ and  $\Omega_\Lambda=0.7$.

\section{The CLASSY sample}\label{sec:data}

The COS Legacy Archive Spectroscopy SurveY (CLASSY) provides high-resolution far-ultraviolet (FUV) Hubble Space Telescope (HST) observations of 45 nearby galaxies \citep{berg2022}. The sample, while low redshift, has higher SFRs than for typical $z\sim0$ galaxies, and is more reminiscent of $z\gtrsim 2$ galaxies. 
The CLASSY dataset consists of spectra providing the complete FUV wavelength coverage of each galaxy, between roughly $1200$\AA\ and $2000$\AA\ rest frame. Observations within CLASSY were taken with the Cosmic Origins Spectrograph (COS) \citep{green2012}. A detailed description of the data reduction is presented in \citet{james2022}. 

\begin{table*}[ht!]
\begin{center}
 \caption{Summary of parameters used in the SALT modeling of UV spectra. }
 \label{tab:pars}
 \begin{tabular}{cccccccccccc}
  \hline
  Parameter & Definition & Parameter Type & Fitting Range\\
  \hline
  $\alpha$ ($^{\circ}$) & Opening angle of the bicone\footnote{A value of $\alpha=90^o$ implies that the outflow is purely spherical, and thus $\psi$ becomes an irrelevant parameter.}  & Free & [0.0, 90.0]\\
  $\psi$ ($^{\circ}$)  & Orientation angle of the bicone & Free & [0.0, 90.0]\\
  $\gamma$ & Wind velocity gradient of the gas, $v(r)\propto r^\gamma$ & Free & [0.5, 2.0]\\
  $\delta$ & Density gradient of the gas, n(r)$\propto r^{-\delta}$ & Semi-Constrained & [$\gamma$-1.5,$\gamma$+1.5]\\
  $\tau_0$ & Optical depth of the gas, where $\tau_0=\tau/(\lambda f_{ul})$ & Free & [0.01, 100.0]\\
  f$_c$ & Covering fraction of the outflowing gas & Free & [0.0, 1.0]\\
  k & Dust opacity & Constrained & \\
  & where $\tau_d$ =
$    \begin{cases}
      \frac{k}{\gamma}ln(\frac{v_w}{v_0})& \text{if} \ \delta=1\\
      \frac{k}{1-\delta}[(\frac{v_w}{v_0})^{(1-\delta)/\gamma}-1] & \text{otherwise.}\\
    \end{cases}  $ &  & \\
  $v_0$ (km s$^{-1}$) & Launch velocity of the gas in the outflow (velocity at $R_{SF}$) & Free &[2.0, 150.0]\\
  $v_w$ (km s$^{-1}$)& Terminal velocity of the gas in the outflow (velocity at $R_w$) & Free & [200.0, 1300.0] \\
  $v_{ap}$ (km s$^{-1}$)& Velocity at the maximum aperture radius (velocity at $R_{AP}$)& Constrained &  \\
  & where $v_{AP}=v_0 \left( \frac{R_{AP}}{R_{50}} \right)^\gamma$ & & \\
\hline
  a  & Depth of the static ISM Gaussian profile & Free & [0.0, 1.0]\\
  $\sigma$ (km s$^{-1}$) & Width of the static ISM Gaussian profile & Free & [0.0, 500.0]\\
  \hline
 \end{tabular}
 \end{center}
\end{table*}

The CLASSY dataset is ideal to study outflows in galaxies \citep{xu2022}. The full available wavelength range includes five resonant transitions of \sitwo\ ($1190.42$\AA, $1193.28$\AA, $1260.42$\AA, $1304.37$\AA, and $1526.72$\AA), \sithree\ ($1206.50$\AA), and two of \sifour\ ($1393.76$\AA \hspace{0.01cm} and $1402.77$\AA). These lines can be used to identify outflows and to compute their properties, including column densities, covering fractions, and velocity fields. The availability of multiple ionization states of the same element allows us to probe gas at different temperatures, and thus fully characterize the multiphase medium \citep[\sitwo\ and \sithree\ are good tracers of gas at around $T=10^4$K, whereas \sifour\ is a tracer of warmer gas of about $10^5$K, see,][]{tumlinson2017}.

As we demonstrate in Section~\ref{sec:comparison}, knowing, and properly accounting for, the spectral resolution of the data is important in deriving the column density of the absorbing material. Because of the size of the COS aperture and the extended nature of the targets, the effective spectral resolution of the COS data varies for each object, as it is the convolution of the intrinsic line spread function of each grating with the galaxy spatial profile in the dispersion direction. Additionally, depending on each galaxy's redshift, the lines of interest fall in either the G130M or G160M grating, which provide different wavelength dispersion (9.97m\AA/px and 12.23m\AA/px, respectively). To estimate the effective resolution for each object, we use the NUV 
half--light radius ($R_{SF}$) as defined and calculated in \citet{xu2022}. For galaxies smaller than the COS aperture, $R_{SF}$ is determined as the radius which contains half of the total light  within the COS acquisition images.
For the (typically more nearby) CLASSY galaxies larger than the COS aperture, we adopt the radius measured from the SDSS $u$-band images. The galaxy acquisition images together with the source profiles in the cross and dispersion directions are  presented in \citet{james2022}. Table~\ref{tab:pars1} in the~\texttt{Appendix} reports the measured $R_{SF}$ for each galaxy.

The galaxy spectra were normalized using stellar template fits obtained as described in \citep[][]{xu2022}. This normalization removes the contribution from stellar wind features, e.g., in the \sifour\ transition, as well as weak and narrow photospheric absorption lines. This normalization step is essential for galaxies containing very massive stars, which, while rare, can substantially influence the shape of the \sifour\ P-Cygni profiles.  This caveat applies in particular to J1200$+$1343 and J1129$+$2034, which were predicted in \citet{martins2023} to host very massive stars.  For a few galaxies, absorption by \hi\ in the \lya\ transition resulted in a poor normalization of some lines of interest (typically the 1206.42\AA\ and 1190.42/1193.38\AA\ lines). In these cases, we  manually re-normalized the spectra using the continuum close to the absorption line.

In the following, we limit the analysis to those galaxies for which the spectra could be realistically modeled and for which lines from all three ionization states of silicon were available. We visually inspected the wavelength ranges of interest to determine if a spectrum was viable for analysis. A spectrum was not modeled if the region of interest is contaminated by either Milky Way absorption, or geocoronal emission. For example, the proximity of the \sithree\ line to the geocoronal \lya\ line occasionally results in no observable P-Cygni profile at $1206.50$\AA. 
Upon further inspection of the spectra, we exclude two additional galaxies for the following reasons.  First, we removed J1525$+$0757 because its spectrum shows two distinct blueshifted absorption components. This galaxy is a clear merging system in the SDSS images. The association between the absorption lines and the galaxy's outflow is therefore questionable, as the gas may appear blushifted as a result of the interaction, rather than an outflow. We also exclude J1416$+$1223, where the relevant silicon lines showed significant redshifted absorption with respect to systemic velocity, more indicative of a galactic inflow than an outflow \citep{rubin2017,carr2022}. This object will be investigated in a forthcoming paper.

Out of the full sample of 45 galaxies, 17 objects have all required absorption lines free from Milky Way absorption and/or geocoronal emission.

\section{SALT Modeling}\label{sec:modeling}
\subsection{Overview of SALT}
The Semi-Analytical Line Transfer (SALT) Model, as outlined in \citet{scarlata2015}, provides a framework for modeling the absorption and scattered emission component of spherical galactic outflows, treated as a series of concentric, expanding shells. 
This model was adapted to account for more general geometries in \citet{carr2018}, generalized velocity and density fields in \citet{carr2023}, and inflows in \citet{carr2022}. The model was successfully used in the study of the properties of outflows in green pea galaxies \citep[see][]{carr2021}. 
 We use this model to extract the physical properties of the outflows in the CLASSY galaxies from the analysis of resonant absorption lines and  fluorescent emission lines of silicon.

SALT assumes that isotropically emitted photons from the stellar population pass through the outflow, where they are resonantly absorbed and scattered. In the most general configuration, the outflow is modeled as a bicone, characterized by its opening angle and orientation with respect to the line of sight. Additionally, SALT allows for velocity and density radial fields (assumed to be power laws with radius), varying covering fraction with radius, and accounts for the presence of dust and spectroscopic aperture. The spectral profile of the resonant line is then dependent on the particular geometry of the outflow, its velocity and density fields, the size of the aperture used for the observations, and the direction through which a galaxy is observed. Depending on the specifics, an outflow can result in a pure absorption profile (e.g., if the bi-cone is oriented parallel to the line of sight), a pure emission profile (e.g., if a bicone is oriented perpendicular to the line of sight), or a combination of the two.
SALT models both the resonant absorption and the re-emission (both resonant and fluorescent) of the photons, thus properly accounting for emission infilling \citep{prochaska2011}. SALT simplifies the radiative transport of resonant photons using the Sobolev approximation \citep{sobolev} and does not account for thermal or turbulent line broadening. The validity of these assumptions has been studied in an extensive comparison using the fully radiative transport code, RASCAS, in \citet{carr2023}.

We have modified SALT to account for the presence of an interstellar medium (ISM). \citet{xu2022} demonstrated the need to account for the presence of absorption in the ISM of some of the CLASSY galaxies, which can be identified at the systemic velocity of each galaxy. We follow \citet{xu2022}, and assume that the ISM absorption component takes the form of a Gaussian centered at the systemic velocity.  This component is removed from the continuum before propagating it through the bi-conical outflow \citep[see][for further clarifications]{xu2022,carr2023}. 
We also allow for the possibility of observing a fluorescent component resulting from the ISM absorption (see Section~\ref{sec:parameters} for details).
We note that the best-fit ISM parameters resulting from our analysis are not expected to match those derived in \citet{xu2022}. This is because SALT consistently accounts for resonant re-emission at systemic velocity (i.e., infilling), while this contribution is neglected in the analysis of  \citet{xu2022}.

Finally, SALT accounts for a limiting spectroscopic aperture which sets the amount of observed emission. This effect is particularly important because the fraction of the remission (of both resonant and fluorescent photons) that are captured in the observed spectrum depends on the relative size of the outflow and the aperture. Not only the equivalent width of the fluorescent emission is impacted by this effect, but also the contribution of line infilling. To model this effect, we use the unvignetted COS aperture, rather than the 1\farcs25 nominal aperture radius which suffers from vignetting. To estimate the unvignetted aperture size we use the radial COS throughput function, which  drops below $80\%$ roughly at 0\farcs7, which we take to be $R_{AP}=$0\farcs7 for this study.\footnote{It should be noted that this estimate of the vignetting radius is based on the instrument optical modeling at Lifetime Position (LP) equal to 1. This is the best that can be done with the available data, but we note that the shape and size of the vignetting will change as a function of position on the COS detector.
}
When multiple transitions of the same ion are available, all lines are modeled together. All SALT parameters describing the outflows are kept consistent across all three absorption lines.

To summarize, SALT uses eight  parameters to describe the outflow, with the possibility  of 2 additional parameters for galaxies hosting a significant ISM component. These parameters are summarized and described in Table~\ref{tab:pars}.
The resonant and fluorescent  transitions that we consider are presented in Table~\ref{tab:atom}.

 \subsection{SALT Limitations}\label{sec:limits}
The SALT model has been previously tested against more complex radiative transport simulations. \cite{carr2018,carr2023} demonstrated that, overall, the accuracy with which the SALT parameters are measured depends on the quality of the data (S/N, resolution, spectroscopic aperture) and on the availability of both absorption and scattered re-emission in the observed spectra. This is because the absorption and emission line profiles are sensitive to different physical components of the outflows. For example, the intensity of the scattered re-emission depends on the opening angle of the outflow, the aperture size, and the overall amount of material. The absorption profile, on the other hand, probes only the absorbing gas along the line of sight. This complexity may introduce degeneracy in some of the parameters, as detailed in \citet{carr2018}, which can be removed with high signal-to-noise and high spectral resolution data. \citet{carr2023} showed that the gradients of the velocity ($\gamma$) and density ($\delta$) fields are typically poorly constrained. Accordingly, we chose to marginalize over these parameters to properly account for their sources of uncertainty. 

However, sources of systematic biases, possibly due to missing physics in the SALT modeling, cannot be accounted for by marginalization.  For instance, because of the Sobolev approximation, SALT assumes that the thermal and turbulent velocities of the outflow are negligible. The impact of this assumption was explored in \citet{carr2023}. The strongest biases are observed in the modeled initial velocities and depend on the gas column densities. For densities lower than $log(N_{True})=16$ the initial velocities are underestimated by approximately 30\%, while for larger column densities the difference can be up to a factor of four. Neglecting the thermal-plus-turbulent velocities also somewhat impacts the densities. For $log(N_{True})>16$, the predicted densities are overestimated by up to a factor of three.
We refer the reader to \citet{carr2018,carr2023} for in depth discussion of these tests.

\begin{table}
\caption{Atomic Data for \sitwo, \sithree, and \sifour \hspace{0.01cm} ions, as well as \stwo\ and \sfour\ for comparison. Data taken from the NIST Atomic Spectra Database$^a$ and \citet{silicon2008}. Example energy level diagrams for such lines of silicon can be found in the Appendix of \citet{carr2021}.}
 \label{tab:atom}
 \centering
 \begin{tabular}{cccccccccccc}
  \hline
  Ion & $\lambda$ (\AA) & Type & $A_{ul}$ $({s^{-1}})$& $f_{ul}$ \\
  \hline
  \sitwo   & 1190.42 & Resonant & $6.53\times10^8$ & 0.277\\
  \sitwo   & 1193.28 & Resonant & $2.69\times10^9$ & 0.575\\
  \sitwo$^*$   & 1194.50 & Fluorescent & $3.45\times10^9$ & 0.737\\
  \sitwo$^*$   & 1197.39 & Fluorescent &$1.40\times10^9$ & 0.150\\
  \sitwo   & 1260.42 & Resonant & $2.57\times10^9$ & 1.22\\
  \sitwo$^*$ & 1265.02 & Fluorescent & $4.73\times10^8$ & 0.113\\
  \sitwo   & 1304.37 & Resonant & $3.64\times10^8$ & $9.28\times10^{-2}$\\
  \sitwo$^*$ & 1309.27 & Fluorescent & $6.23\times10^8$ &$8.00\times10^{-2}$ \\
  \sitwo   & 1526.72 & Resonant & $3.81\times10^8$ & 0.133\\
  \sitwo$^*$ & 1533.45 & Fluorescent & $7.52\times10^8$ & 0.133 \\
  \sithree            & 1206.50 & Resonant & $2.55\times10^9$ & 1.67\\
  \sifour                & 1393.76 & Resonant & $8.80\times10^8$ & 0.513\\
  \sifour                & 1402.77 & Resonant & $8.63\times10^8$ & 0.255\\
  \hline
    \stwo            & 1253.81 & Resonant & $5.12\times10^7$ & $1.21\times10^{-2}$\\
    \sfour  & 1062.65 & Resonant & $1.48\times10^8$ & $5.00\times10^{-2}$\\
  \hline
  
 \end{tabular}
 \newline
 $^a$ http://www.nist.gov/pml/data/asd.cfm
\end{table}

\subsection{Estimate of Outflow Parameters}\label{sec:parameters}

To estimate the properties of the  outflows in the CLASSY galaxies, we fit their UV spectra with SALT.  As discussed in Section~\ref{sec:data}, the effective spectral resolution is object dependent, and this effect needs to be accounted for in the fitting procedure. We convolve the SALT-generated spectra with a Gaussian kernel before they are compared to the data. 
The standard deviation of this Gaussian is chosen to be the physical size of the galaxy converted to \AA\ using the pixel scale of the grating in which the line is observed. Specifically, we use 22.9m$''$/px and 24.2m$''$/px for lines in the G130M and G160M grating, respectively.

\begin{figure*}[ht!]
    \centering
    \includegraphics[width=0.9\linewidth]{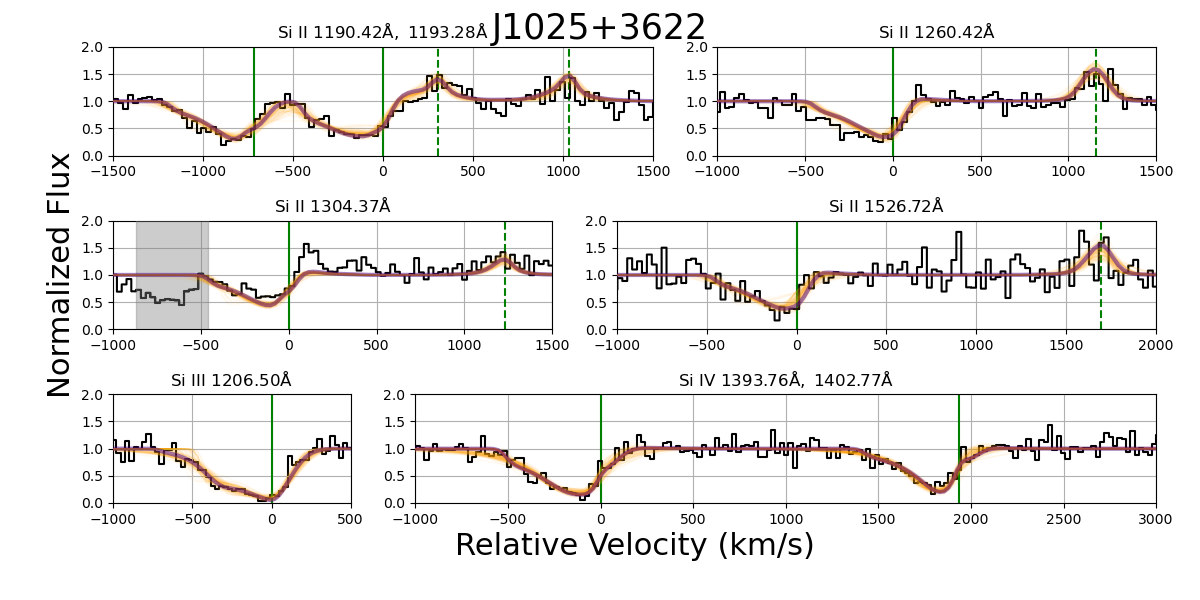}
    \caption{Example SALT fits for CLASSY galaxy J1025+3622. Each panel shows the best-fit SALT model in red, together with 100 models chosen randomly from the posterior distributions in orange. The green solid lines show the positions of resonant transitions, while the dashed green lines show the positions of the fluorescent emission. The gray shading indicates a masked region. This example shows how SALT can accurately reproduce both the absorption and fluorescent emission in a consistent way, see e.g., the \sitwo\ modeled profiles in the top two rows. The complete figure set (17 images) is available in the online journal.}
    \label{fig:exa}
\end{figure*}

Before fitting the model to the data, we also mask regions of the spectra contaminated by Milky Way absorption and geocoronal emission. Specifically, we mask 125 km s$^{-1}$ around the central wavelengths of the  spectral features reported  in \texttt{Table \ref{tab:mw}}.  
We note that blending of metal lines may be an issue for some of the transitions. For example, there is an Fe\thinspace{\sc iii} line at $1260.37$\AA, which could potentially contaminate the $1260.42$\AA\ \sitwo\ line. This line, however, is known to be weak \citep[see][]{ekberg1993}, and we assume it to be negligible. Additionally, the \sitwo\ $1304.37$\AA\ transition may be compromised due to the O\thinspace{\sc i}$^*$ $1304.86$ and $1306.03$\AA\ fluorescent channels associated with the O\thinspace{\sc i} $1302.17$\AA\ resonant line. While often ignored, possible infilling may affect the modeling of this line; therefore we never fit the \sitwo~$1304.37$\AA\ line on its own; it must be fit in conjunction with (at minimum) either the more optically thick $1190.42/1193.28$\AA\ or $1260.42$\AA\ \sitwo\ lines. An example of this infilling can be seen in the galaxy J1025$+$3622, shown in Figure~\ref{fig:exa}. Finally, an additional narrower mask of 50 km s$^{-1}$ is placed at the \stwo\ line of 1259.52\AA\ in the observed galaxy's rest frame, as this may overlap with the stronger \sitwo\ 1260.42\AA\ absorption line. 
The \stwo\ transition is weak, and therefore masking out the small velocity range gets the strongest absorption while leaving the 1260.42\AA\ line intact.

\begin{table}
 \caption{Milky Way and Geocoronal Lines}
 \label{tab:mw}
 \centering
 \begin{tabular}{cccccccccccc}
  \hline
  Ion & Wavelength (\AA)\\
  \hline
  \hi                & 1215.67\\
  \ci               & 1277.21 \\
  \cii & 1334.53, 1335.71\\
  \nwon                 & 1199.55, 1200.22, 1200.71\\
  \nv                & 1235.81, 1242.80\\
  \oi  & 1302.17\\
  \sitwo   & 1190.42, 1193.28, 1260.42, 1304.37, 1526.72\\
  \sithree            & 1206.50\\
  \sifour                & 1393.76, 1402.77\\
  \stwo & 1250.59, 1253.81, 1259.52\\
  \hline
 \end{tabular}
\end{table}

When modeling transitions with fluorescent channels, we consider two possibilities for the ISM component: (1): only ISM resonant absorption, and (2): ISM resonant absorption $+$  fluorescent emission. The second option is motivated by the idea that the fluorescent emission from the ISM is produced within the galaxy, rather than tracing the outflow, and is therefore fully captured by the spectroscopic aperture. 
We choose which model best fits the data according to the Bayesian Information Criterion \citep[BIC,][]{schwarz1978}. 

We use the \texttt{emcee} Markov Chain Monte-Carlo (MCMC) sampler \citep{foreman2013} to compute the posterior distribution of the outflow and ISM parameters, assuming uniform priors in the ranges provided in the last column of \texttt{Table \ref{tab:pars}}. We refrain, however, from letting $\gamma$ and $\delta$ be entirely free by noting that the mass outflow rate should not deviate substantially from the constant mass outflow rate (i.e., $\delta=\gamma+2$ \citealt{carr2018}). Therefore, we constrain $\delta$ to be within a value of $\pm$1.5 from the constant mass outflow rate prediction, i.e.  $\gamma+0.5\leq \delta \leq \gamma+3.5$. This range includes the median $\delta$ value found by \cite{chisholm2017}, and still allows for deviations from mass conservation due to, for example, a change in the ionization state \citep{werk2014}. The fitting range for $\gamma$ is chosen to be bounded on the lower end by 0.5 as when $\gamma$ approaches 0, the Sobolev approximation assumption breaks down \citep[for further details, see Equation 24 in][]{carr2023}.

We constrain the dust opacity, $k$, using the measured color excess \citep[taken from][]{berg2022} and assuming a Calzetti attenuation law \citep{calzetti1997} to compute $A_{UV}$. We convert $A_{UV}$ to the dust optical depth with $A_{UV} \approx 1.086 \tau_{UV}$.

\figsetstart
\figsetnum{1}

\figsetgrpstart
\figsetgrpnum{1.1}
\figsetgrptitle{J0021+0052}
\figsetplot{J0021+0052.png}
\figsetgrpnote{SALT fits for J0021+0052. The \sitwo\ 1190.42\AA/1193.28\AA\ doublet was not fit due to lack of data in the region of interest.}
\figsetgrpend

\figsetgrpstart
\figsetgrpnum{1.2}
\figsetgrptitle{J0036-3333}
\figsetplot{J0036-3333.png}
\figsetgrpnote{SALT fits for J0036-3333. The \sitwo\ $1190.42$\AA/$1193.28$\AA\ doublet was not fit due to geocoronal Lyman $\alpha$ contamination.}
\figsetgrpend

\figsetgrpstart
\figsetgrpnum{1.3}
\figsetgrptitle{J0405-3648}
\figsetplot{J0405-3648.png}
\figsetgrpnote{SALT fits for J0405-3648. The \sitwo\ $1190.42$\AA/$1193.28$\AA\ doublet was not fit due to Milky Way contamination.}
\figsetgrpend

\figsetgrpstart
\figsetgrpnum{1.4}
\figsetgrptitle{J0926+4427}
\figsetplot{J0926+4427.png}
\figsetgrpnote{SALT fits for J0926+4427.}
\figsetgrpend

\figsetgrpstart
\figsetgrpnum{1.5}
\figsetgrptitle{J0934+5514}
\figsetplot{J0934+5514.png}
\figsetgrpnote{SALT fits for J0934+5514. The \sitwo\ $1190.42$\AA/$1193.28$\AA\ doublet was not fit due to Milky Way contamination.}
\figsetgrpend

\figsetgrpstart
\figsetgrpnum{1.6}
\figsetgrptitle{J0938+5428}
\figsetplot{SALT fits for J0938+5428.png}
\figsetgrpnote{SALT fits for J0938+5428.}
\figsetgrpend

\figsetgrpstart
\figsetgrpnum{1.7}
\figsetgrptitle{J0940+2935}
\figsetplot{J0940+2935.png}
\figsetgrpnote{SALT fits for J0940+2935.}
\figsetgrpend

\figsetgrpstart
\figsetgrpnum{1.8}
\figsetgrptitle{J1024+0524}
\figsetplot{J1024+0524.png}
\figsetgrpnote{SALT fits for J1024+0524. The \sitwo\ $1260.42$\AA\ line was not fit due to Milky Way contamination.}
\figsetgrpend

\figsetgrpstart
\figsetgrpnum{1.9}
\figsetgrptitle{J1025+3622}
\figsetplot{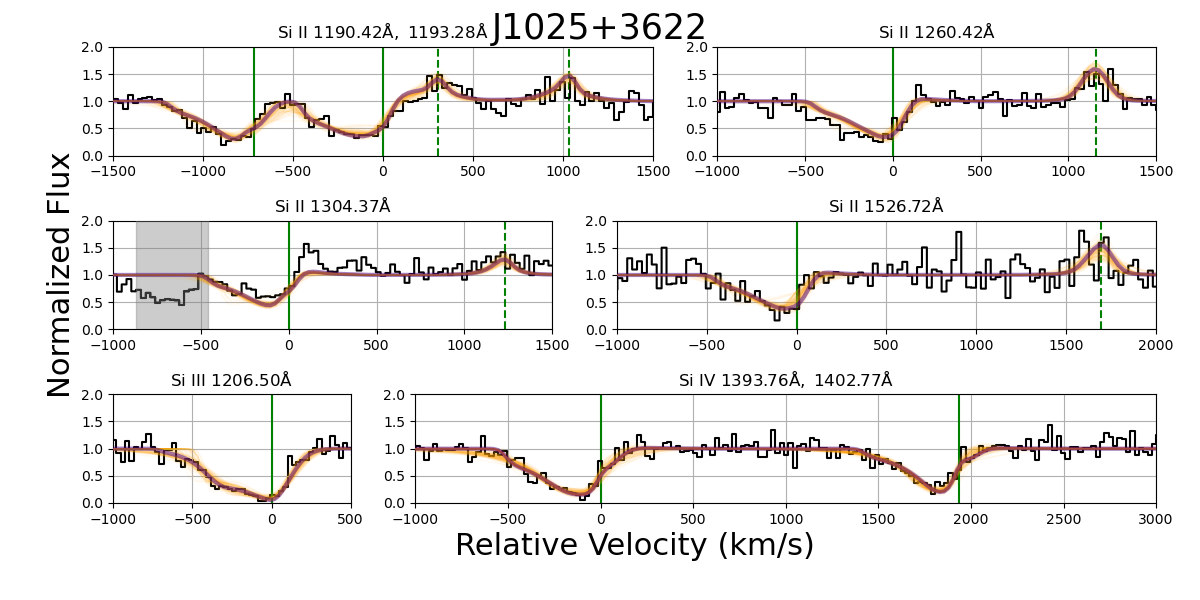}
\figsetgrpnote{SALT fits for J1025+3622.}
\figsetgrpend

\figsetgrpstart
\figsetgrpnum{1.10}
\figsetgrptitle{J1105+4444}
\figsetplot{J1105+4444.png}
\figsetgrpnote{SALT fits for J1105+4444. The \sitwo\ $1190.42$\AA/$1193.28$\AA\ doublet was not fit due to lack of data in the region of interest}
\figsetgrpend

\figsetgrpstart
\figsetgrpnum{1.11}
\figsetgrptitle{J1112+5503}
\figsetplot{J1112+5503.png}
\figsetgrpnote{SALT fits for J1112+5503.}
\figsetgrpend

\figsetgrpstart
\figsetgrpnum{1.12}
\figsetgrptitle{J1150+1501}
\figsetplot{J1150+1501.png}
\figsetgrpnote{SALT fits for J1150+1501. The \sitwo\ $1190.42$\AA/$1193.28$\AA\ doublet was not fit due to Milky Way contamination.}
\figsetgrpend

\figsetgrpstart
\figsetgrpnum{1.13}
\figsetgrptitle{J1200+1343}
\figsetplot{J1200+1343.png}
\figsetgrpnote{SALT fits for J1200+1343.}
\figsetgrpend

\figsetgrpstart
\figsetgrpnum{1.14}
\figsetgrptitle{J1314+3452}
\figsetplot{J1314+3452.png}
\figsetgrpnote{SALT fits for J1314+3452. The \sitwo\ $1190.42$\AA/$1193.28$\AA\ doublet was not fit due to Milky Way contamination.}
\figsetgrpend

\figsetgrpstart
\figsetgrpnum{1.15}
\figsetgrptitle{J1359+5726}
\figsetplot{J1359+5726.png}
\figsetgrpnote{SALT fits for J1359+5726. The \sitwo\ $1260.42$\AA\ line was not fit due to the presence of the geocoronal Lyman $\alpha$ line.}
\figsetgrpend

\figsetgrpstart
\figsetgrpnum{1.16}
\figsetgrptitle{J1428+1653}
\figsetplot{J1428+1653.png}
\figsetgrpnote{SALT fits for J1428+1653.}
\figsetgrpend

\figsetgrpstart
\figsetgrpnum{1.17}
\figsetgrptitle{J1429+0643}
\figsetplot{J1429+0643.png}
\figsetgrpnote{SALT fits for J1429+0643.}
\figsetgrpend

\figsetend

Example fits for each of the silicon lines for the CLASSY galaxies are reported in \texttt{Figure \ref{fig:exa}} and we also show an example corner plot for one galaxy Appendix~\ref{app:corner}. Additionally, Table~\ref{tab:pars1} in Appendix~\ref{app:fits} reports the best-fit parameters of the outflows for all ionic states of silicon. As explained in Section~\ref{sec:limits} $\delta$ and $\gamma$ are less constrained observationally relative to other SALT parameters. Accordingly, we marginalize over their values in the estimate of the best fit SALT parameters, which are computed as the median of the posterior distribution. For each parameter we report the associated credibility interval corresponding to the range between the 16$^{th}$ and 84$^{th}$ percentiles of the posterior distributions.

\begin{figure*}[ht]
    \centering
    \includegraphics[width=1.\linewidth]{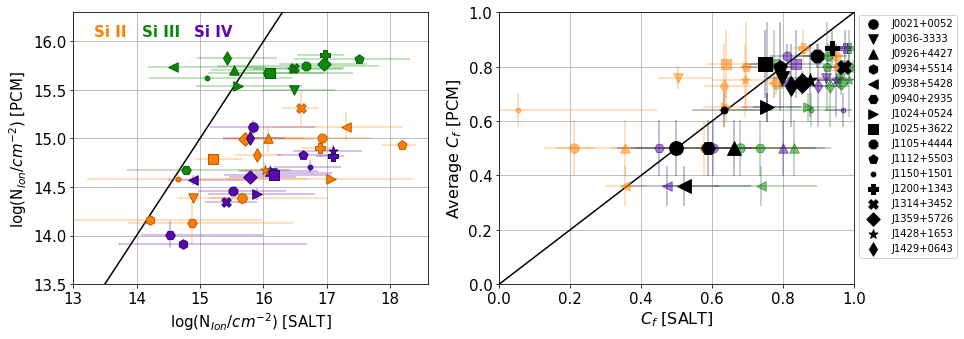}
    \caption{\textbf{Left}: Comparison between column densities derived using SALT and the Partial Covering Fraction method (PCM) from \citet{xu2022}. We report column densities for all three ionization states of silicon. The  PCM \sitwo\ and \sifour\ column densities were computed directly from the data, while the \sithree\ values were obtained using photo-ionization modeling. 
    \textbf{Right}: Comparison between the individual ion covering fractions computed with SALT and the average covering fraction computed using the PCM method. The two methods on average are positively correlated, with the black points representing the average SALT result across all ionization states.}
    \label{fig:3comp}
\end{figure*}

\section{Derived quantities}
\label{sec:quantity_definitions}
We use the best-fit parameters described above to compute a number of physical properties for the outflows, reported in Table~\ref{tab:pars1}. 

The physical quantities we consider are the number density $n_0$, the terminal outflow radius $R_W$, the column density $N_{ion}$, and the outflow rate ($\dot{M}_{\rm ion}$). We compute these quantities for each silicon ionization state, as follows: 

\begin{equation}\label{eq:n0}
n_0=\frac{m c \tau v_0}{\pi e^2 R_{SF}}, 
\end{equation}

\begin{equation}\label{eq:rw}
R_w=R_{SF}\left(\frac{v_w}{v_0}\right)^{1/\gamma},
\end{equation}

An extended source allows for variations in the column density along different lines of sight. To account for this, we compute an 'effective' column density by uniformly sampling column density measurements across the surface of the source and taking an average.  Instructions on how to compute the column density, $N_{l,\xi}$, in the SALT formalism, along an arbitrary sight line (designated by coordinates $(l,\xi)$) are provided in Appendix~\ref{app:column_density}.  In the analysis that follows, by `column density' we refer to the quantity,
\begin{eqnarray}\label{eq:NN}
    N_{ion} = \sum_i^P\frac{N_{l(i),\xi(i)}}{P},
\end{eqnarray}
where the sum is taken over $P$ arbitrary lines of sight. 
We find that our results converge to roughly two significant figures by $P=10^4$.

For each ionization state of silicon, the mass outflow rate as a function of distance, $r$, from the galaxy, i.e., the mass passing through a radius $r$ of the outflow per second, is computed as:

\begin{equation}
\label{eq:Mass}
\dot{M}_{\rm ion} (r)=\Omega f_c m_{Si}n_{0} v_0 R_{SF}^2 \left(\frac{r}{R_{SF}}\right)^{(2+\gamma-\delta)},
\end{equation}

\noindent 
where $m_{Si}$ is the mass of silicon, $\Omega=4 \pi (1-\cos{\alpha})$, and all parameters correspond to those of the specific ionization state. 
Note that, in general, the mass outflow rate is a function of radius, $r$. Observationally, this is difficult to constrain without a full model of the outflow, as each observed spectral bin ($v_{obs}$) receives contributions from gas that is located at different radii and different velocities, due to projection effects.
We choose, however, to compute the mass outflow rate at $r=R_{SF}$ to better compare with the SFR.  Computing the  outflow rates at larger radii could introduce a lag between SFR measurements and the perceived mass outflow rate values which would jeopardize the significance of the mass loading factors.

We compute the total mass outflow rate of silicon assuming that silicon only exists in the three observed ionization states  \citep[as in][]{chisholm2018,mckinney2019}; thus the total silicon mass outflow rate can be determined by summing the three individual components, i.e., $\dot{M}_{\rm Si} = \dot{M}_{\rm Si\,II} + \dot{M}_{\rm Si\,III} + \dot{M}_{\rm Si\,IV} $. 
Classically, photoionization models would show if \sitwo, \sithree, and \sifour\ contain the majority of the gas. However, this modeling assumes that all of the gas is co-spatial and excited by the same mechanism. In reality, these three ionization states trace very different gas temperatures \citep[see][]{tumlinson2017} and can be associated with different components of the CGM where the gas can be photoionzed (as assumed in photoionization models) or, for example, heated by outflow-driven shocks. As an example, if the cooler outflow was immersed in a hotter, volume filling (\sifive\ or higher) component, we would not be able to infer this from applying a ``correction" to the cooler clumps. Photoionization modeling would just tell us how much \sifive\ is co-spatial with the lower ionization states.

Finally, from $\dot{M}_{Si}$ we compute the total hydrogen mass outflow rate for each galaxy, assuming that the outflow has the same metallicity as that derived using the nebular lines from the SF regions:

\begin{equation}\label{eq:hydromor}
\dot M_{H}=\dot M_{Si}\frac{1/Z_{Galaxy}}{Z_{Si,\odot}},
\end{equation}

\noindent
where $Z_{Galaxy}$ is the ratio of metals to hydrogen and has units of the solar metallicity and $Z_{Si,\odot}=7.167\times 10^{-4}$ is the silicon mass fraction for the Sun, taken from \citet{solar}. 
Similarly, the total  hydrogen column density can be computed given the individual silicon ions' column density, using the solar silicon number fraction:

\begin{equation}\label{eq:nhtotal}
N_{H}=\frac{1}{Z_{Galaxy}}\left[\frac{\rm H}{\rm Si}\right]_\odot \sum_{ion} N_{ion}.
\end{equation}

\section{Comparison to the PCM Method}
\label{sec:comparison}

As mentioned in the introduction, the classical approach to estimate the column density and in general the properties of outflowing gas identified in galaxies is to use the partial covering fraction method. 
The outflow properties of the CLASSY galaxies were studied using this approach in \citet{xu2022}. It is instructive to compare the results of the classical method with those obtained using the semi-analytical model described above; to understand how robust and model dependent our knowledge of galaxy outflows is.  

The presence of a  ISM absorption at the systemic velocity of the galaxy complicates the identification of outflowing gas along the line of sight. \citet{xu2022} conduct an `F-test' to determine whether an outflow model was an improvement over a purely static ISM model, i.e., well fit by an absorption line centered at the systemic velocity.  Similarly, as explained in Section~\ref{sec:parameters}, we use the BIC value to identify the presence of outflows.
In Table~\ref{tab:pars1} a dagger identifies galaxies for which the F-test in \citet{xu2022} does not support the presence of an outflow. Table~\ref{tab:pars1} shows that the two analyses disagree for one galaxy, J0405$-$3648, where the SALT modeling suggests the presence of an outflow, while PCM does not. The best fit for this galaxy is shown in Figure Set~\ref{fig:exa}. The \sithree\ and \sifour\ profiles are clearly blueshifted with respect to the systemic velocity, indicating the presence of an outflow in these ionization states of silicon. The \sitwo\ profiles are less convincing.

In what follows we proceed with the comparison of the outflow properties for the 16 (out of 17) CLASSY galaxies for which both analyses agree on the detection of an outflow.

\subsection{Column Densities and Covering Fraction}

\citet{xu2022} used the PCM method to derive the column density and covering fraction of the absorbing material as a function of projected velocity. As in this study, they modeled multiple transitions in different ionization states of silicon. 
To be able to remove the degeneracy between optical depth and covering fraction, the PCM requires the availability of at least two transitions from the same ion. This is because in the presence of partial covering, $\frac{I}{I_0}$ will not go to zero when saturated.
If all lines are optically thin, their strength is proportional to $f_{ul} \lambda$ (where $f_{ul}$ is the oscillator strength of the transition), whereas if they are all optically thick, they will have the same strength. 
Because only one line of \sithree\ is available in the observed wavelength range, \citet{xu2022} do not measure the \sithree\ column density directly. Instead, they use photo-ionization modeling to reproduce the measured \sitwo/\siiv\ line ratios and predict the contribution of \sithree\ to the total Si.

In the SALT forward-modeling approach of the absorption line profiles, we are able to remove the degeneracy between covering fraction and optical depth even when only one resonant transition is available. This is because SALT treats each spectral point in the observed velocity space independently. In this space, 
the optically thick and thin components of the outflowing gas dominate in different velocity ranges. SALT uses this full spectral information to constrain the outflow  parameters such as $\tau_0$ and $f_c$, breaking the degeneracy that hampers the single-line PCM modeling. In other words, by modeling the full shape of the profile,  SALT is able to fit to the optically thin portions of the spectrum and extrapolate the assumed analytical forms into the optically thick part of the line profile.

In Figure \ref{fig:3comp} (left panel) we compare the logarithm of the column densities computed using the PCM and SALT methods for the three ionization states of silicon. This comparison shows that the two methods can differ substantially. Focusing first on \sitwo\ and \sifour\ ionization states, as they are modeled directly from the data in both analysis, we can see that overall there is a broad correlation between the two methods, but the slope is much shallower than one ($0.23\pm0.04$)\footnote{A slope of one would imply a simple scaling factor between the column densities obtained with the two methods.}, indicating that the difference between the two methods is a function of the column density itself. Specifically, larger column densities are associated with larger differences,  reaching  to two orders of magnitude for the largest values.  The comparison between the \sithree\ column densities  shows that there is no correlation between the two  methods (the slope is $0.05\pm0.04$, consistent with zero). This result is probably a consequence of the fact that in \citet{xu2022} the \sithree\ column densities are not derived from the data directly, but from the use of photo-ionization models tuned to reproduce the \sitwo/\sifour\ ratios.

Figure~\ref{fig:3comp} also shows that the reported uncertainties for the SALT column densities are much larger than the uncertainties computed from the PCM analysis, for all ionization states of silicon. While the PCM errorbars account for the uncertainty in the data, they do not attempt to quantify the contribution to the uncertainty resulting from the unknown underlying gas density and velocity distribution, or from the photoionization modeling used to compute the \sithree\ values. The geometrical effects, instead, are included in the SALT estimates of the error bars, as our posterior distributions are properly marginalized to account for the uncertainty contribution of all parameters. We believe our error estimates are more realistic. 

In Figure \ref{fig:3comp} (right panel) we compare the covering fractions computed with SALT to the average PCM covering fractions reported in \citet{xu2022}. In the SALT formalism, $f_c$ represents the fraction of a bi-conical shell covered by material (see \citet{carr2018} for a definition), and is not equal to the traditional definition of the covering fraction assumed by \cite{xu2022}.  We compute an equivalent value in the SALT formalism by using the same Monte Carlo approach used to compute the column density in Section~\ref{sec:quantity_definitions}.  Specifically, we generate a covering fraction, $C_f$, by registering the detection of material along lines of sight with non-zero column densities and scale by $f_c$ to account for holes in the bi-conical outflow. For SALT, we report both the covering fraction computed for each individual ion, and the average of the three. We find that SALT's covering fractions are broadly consistent with those derived from the PCM method, albeit with a substantial scatter. Both models predict large covering fractions in most cases (rarely below $50\%$) and SALT tends to predict a larger $C_f$ than the PCM, in particular for \sithree\ and \sifour. SALT \sitwo\ $C_f$, on the other hand, appear to be systematically lower than the average PCM values. As we discussed before, covering fraction and column density are degenerate, in the sense that the same residual flux could be reproduced with either a large covering fraction and low column density, or a small covering fraction and large column density. The result that the column densities are larger in the SALT analysis compared to the PCM method, while the covering fractions broadly agree, suggests that the column density differences are not due to this degeneracy. 

\begin{figure*}[!htb]
    \centering
    \includegraphics[width=1.\linewidth]{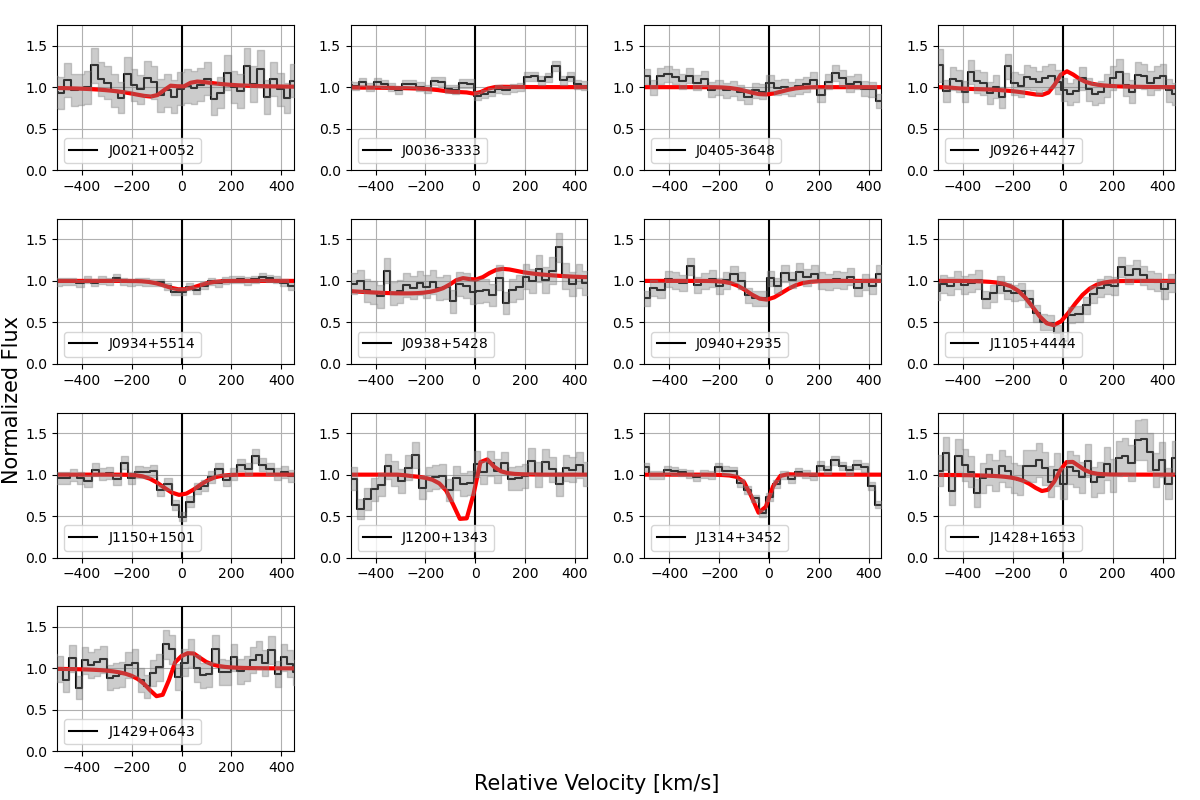}
    \caption{Comparison between the predicted \stwo~1254\AA\ line profile (red line) and the observed data (black line). The predicted profiles are computed using the best-fit geometrical parameters derived from the \sitwo\ fits,  scaled to the oscillator strength of the line and solar ratio of sulfur to silicon for galaxies with  non-contaminated \stwo\ line.}
    \label{fig:sulfur2}
\end{figure*}

\begin{figure*}[!htb]
    \centering
    \includegraphics[width=1.\linewidth]{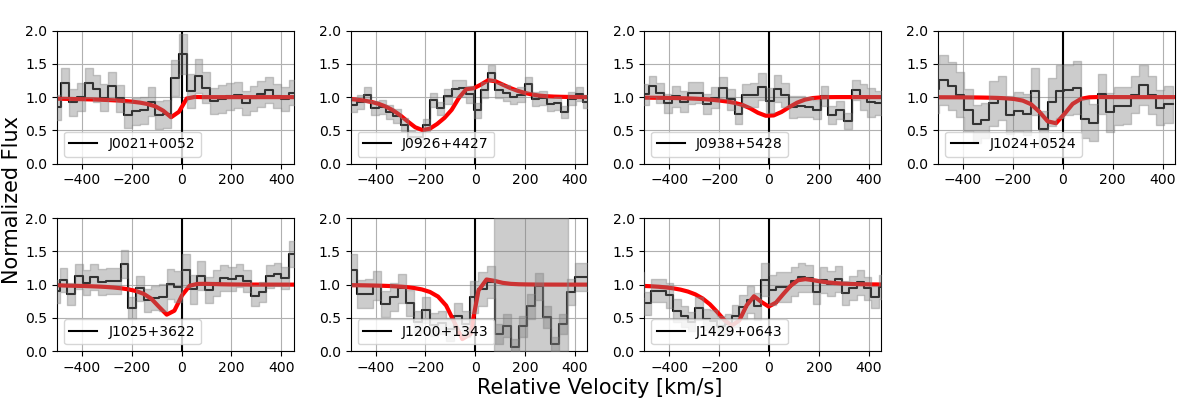}
    \caption{Same as Figure~\ref{fig:sulfur4} but for \sfour~1063\AA, using the \sifour\ fits.}
    \label{fig:sulfur4}
\end{figure*}

\subsection{Comparison to Sulfur Lines}
The comparison of the column densities of silicon shows that in many cases, the full SALT modeling results in values for $N_{\rm SiII}$ and $N_{\rm SiIV}$  that are one to two orders of magnitude larger than the PCM values. To explore the origin of this disagreement, we start by performing a sanity check using absorption lines of sulfur.  Using stacked spectra, \citet{xu2022} showed that the \sitwo\ and \sifour\ columns derived from the PCM are in excellent agreement with the respective column densities from the much weaker \stwo\ and \sfour\ lines (assuming solar S/Si relative abundance). Here we compare the SALT predictions of the sulfur lines for individual spectra. 

For this test, we used the modeling results for the silicon lines to predict the corresponding sulfur absorption line profiles. To compute the sulfur lines, we keep the geometrical parameters fixed to those obtained from the silicon lines, while changing the atomic parameters to reflect the specifics of the sulfur transitions, and scaling the absorption profiles by the relative solar abundance of S/Si. We use the \sitwo\ and \sifour\ parameters to predict the \stwo~1253\AA\ and \sfour~1063\AA\ line profiles. When predicting the sulfur lines we do apply a correction for the ionization fraction, using the results of the  photoionization calculations of \citet{xu2022}. Specifically, we use ICF(\stwo)/ICF(\sitwo)$\sim0.6$, and ICF(\sfour)/ICF(\sifour)$\sim0.8$.

Figures \ref{fig:sulfur2} and \ref{fig:sulfur4} show the comparison between the predicted sulfur line profiles and the spectra for the galaxies with uncontaminated lines. This comparison demonstrates that the silicon-based SALT models do not predict unphysically strong absorptions for the sulfur lines. This is true for both the \stwo\ and the \sfour\ ionization states. We note that the \stwo\ resonant transition has no fluorescent channel, so it is strongly affected by in-filling due to resonant photons scattering back toward the observer. This infilling is properly accounted for in the SALT modeling (including the effect of spectroscopic aperture).  Given the simplicity of the approach, it is in fact remarkable that for some of the galaxies the agreement between predicted and observed is so good (e.g., the \stwo\ in J1314$+$3452). As an example, the SALT \sifour\ column density for  J1200$+$1343 is two orders of magnitude larger than the one computed by the PCM method. Yet, the prediction of the \sfour\ profile shown in Figure~\ref{fig:sulfur4} shows signs of under-predicting the actual absorption.

\subsection{Discussion of HI Column Density}

Low ionization absorption lines have often been used in the literature to quantify the column density of neutral hydrogen \citep[e.g.,][]{shapley2003,saldana2022}.
In particular \sitwo\ is believed to trace \hi, given its low ionization potential. \cite{xu2022}, however, find that typically only 1–10\% of the \sitwo\ gas detected in the CLASSY galaxies is associated with \hi\ gas. The column density of neutral hydrogen along the line of sight can be estimated directly via the detection of absorption lines of the Lyman series. The \lya\ absorption profile, for example, was used in \citet{hu2023} to constrain the column density of \hi\ Damped \lya\ absorbers (DLAs) in the CLASSY galaxies. 

We scale the \sitwo\ column density to the column density of \hi, in the manner discussed in Section~\ref{sec:quantity_definitions}. At first glance, it seems natural to directly compare the \hi\ results of SALT and PCM to the direct measurement of \hi\ in the DLAs. However, both the SALT and PCM model only model the \hi\ in the outflow, whereas the DLA includes \hi\ contribution from both the ISM and the outflow. Accordingly, both SALT and PCM column densities are likely an underestimate of the total \hi\ column density, but the extent to which this is an underestimate is uncertain.

\begin{figure}[ht]
    \centering
    \includegraphics[width=1.\linewidth]{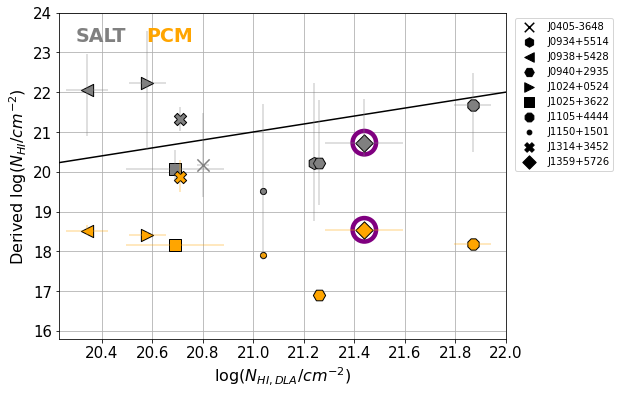}
    \caption{Comparison of HI derived from three different methods: The DLA HI column density \citep{hu2023}, HI derived from the PMC \citep{xu2022}, and HI derived from SALT. SALT agrees closer in magnitude with the DLA result than it does with the PCM result.}
    \label{fig:HIcomp}
\end{figure}

Figure~\ref{fig:HIcomp} shows that the SALT \hi\ column densities are below the DLA values in 7 out of 10 galaxies, with the remaining three objects having a larger column density compared to the DLA values. On the other hand, the \hi\ column densities derived from the photoionization modeling of the silicon lines  are between two and three orders of magnitude the densities of the DLA. Note that scaling the PCM \sitwo\ column densities (instead of using the photoionization modeling) would not change the results. The difference between the PCM and SALT estimate of the \hi\ column densities is not surprising as it reflects the difference between the \sitwo\ column densities already discussed in Section~\ref{sec:comparison}. 

However as discussed above, this comparison is uncertain as the DLA estimates include the contribution of H in the ISM. It is therefore instructive to only look at case studies in which the gas profiles are dominated by the outflowing component. Take J1359$+$5726, for which the \sitwo, \sithree, and \sifour\ are clearly dominated by the outflow component rather than the static ISM, in both the PCM and SALT modeling. This galaxy is highlighted in Figure~\ref{fig:HIcomp} with a purple circle. The DLA \hi\ column density is $\log{N_{HI}}=21.55\pm0.15$~cm$^{-2}$, about a factor of 7 larger than  the  corresponding SALT value, $\log{N_{HI}}=20.73^{+1.07} _{-0.33}$cm$^{-2}$. On the other hand, the PCM \hi\ column density from the cloudy modeling of this galaxy is $18.54^{+0.03} _{-0.03}$~cm$^{-2}$, three orders of magnitude lower than the DLA estimate. In this specific case, reconciling the PCM value with the DLA \hi\ value would imply that most of the \hi\ detected in the DLA has no silicon associated to it, or that most of the silicon is in an ionization state other than those covered by the CLASSY data, which seems unlikely.

\begin{figure*}[ht]
    \centering
    \includegraphics[width=1.\linewidth]{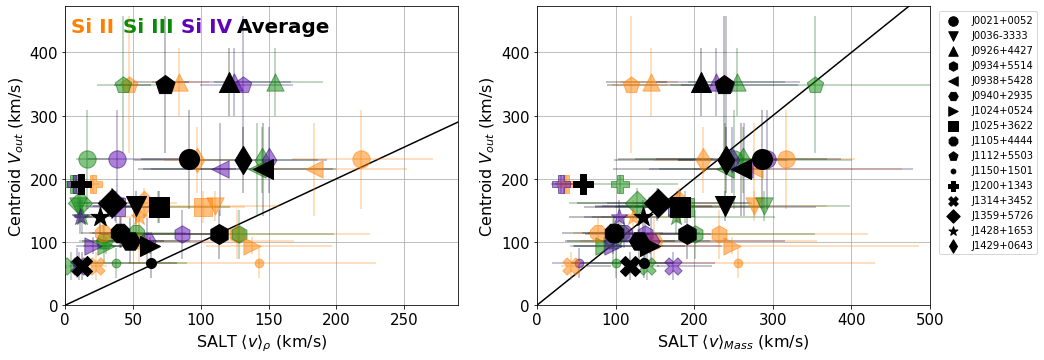}
    \caption{\textbf{Left:} Comparison between $V_{out}$ using the Partial Covering Fraction method (PCM) from \citet{xu2022} and a density weighted average velocity as derived by SALT. \textbf{Right:} Comparison between $V_{out}$ using the PCM from \citet{xu2022} and a mass weighted average velocity as derived by SALT. We report the individual ionization SALT velocities as well as the average ionization state SALT velocity in black. Both the mass and density weights are positively correlated with the PCM results, but it is the mass weighted velocity that agrees most prominently with the PCM results. The density weighted velocity is lower for SALT than for PCM, based on the fact that the power laws for density indicate a substantial component of material is contained within small radii of the outflow.}
    \label{fig:voutvave}
\end{figure*}

\subsection{Velocity Comparison}\label{sec:velocity}

It is instructive to estimate a bulk velocity for the identified outflows, and compare them with traditional estimates based on the position of maximum absorption  \citep[e.g,][]{Rivera2015}. For SALT, we compute the density weighted average velocity as
\begin{equation}
\langle v\rangle _{\rho}=\frac{\int^{R_W}_{R_{SF}} v(r)n(r)dr}{\int^{R_W}_{R_{SF}} n(r)dr},
\label{eq:vel}
\end{equation}
and the mass weighted average velocity as:
\begin{equation}
\langle v\rangle _{Mass}=\frac{\int^{R_W}_{R_{SF}} v(r)n(r)r^2dr}{\int^{R_W}_{R_{SF}} n(r)r^2dr}.
\label{eq:velm}
\end{equation}

The density-weighted velocity corresponds to the velocity of the gas that most contributes  to the observed absorption in the spectra. The mass-weighted velocity, on the other hand, provides an estimate of the velocity at which the bulk of the material is moving \citep[e.g.,][]{schneider2020}. 
We compute these velocities for each ionization state of silicon as well as the mean of the three ionization states. In Figure~\ref{fig:voutvave} we compare the average velocities derived with Equations~\ref{eq:vel} and ~\ref{eq:velm} to the outflow centroid velocities computed in \citet{xu2022}. \citet{xu2022} report the median value of central velocities from all troughs that have passed the F-test (including the fore-mentioned Si lines, as well as \oi~1302 and \cii~1334), commenting that there is strong correlation between the velocities derived from \sitwo~1260 and \sifour~1393, as well as those derived from different transitions of the same ion. The comparison shows that, regardless of the ionization state considered, the outflow velocities derived in \citet{xu2022} are systematically higher (shifted on average by 100 km/s) than the density-weighted velocities computed using the SALT velocity and density profiles, and are more in agreement with the mass-weighted values.
The difference between the density-weighted and centroid velocities likely results from the fact that SALT accounts for infilling of the absorption lines \citep{prochaska2011,scarlata2015}, and, to some extent, the spectral resolution, particularly in the galaxies with strong ISM absorption.

\begin{figure}[ht]
    \centering
    \includegraphics[width=1.\linewidth]{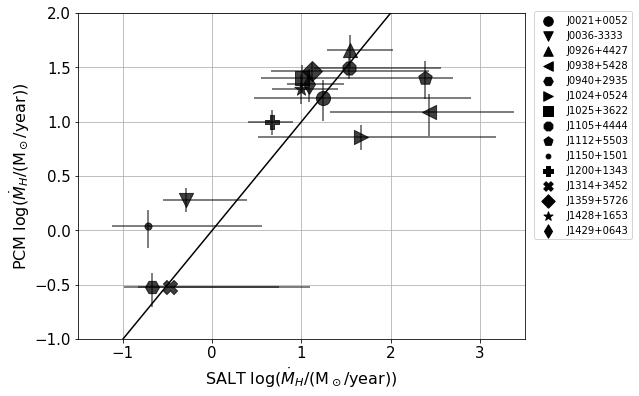}
    \caption{Comparison between PCM and SALT derived total hydrogen mass outflow rates. The two methods have a strong positive correlation with each other, with the galaxies clustering near the black one-to-one line.}
    \label{fig:mhcomp}
\end{figure}

\subsection{Total Mass Outflow Rate Comparison}

With the column density of silicon and the velocity field computed in the previous Sections, we are in the position to compute the total mass outflow rates.  In Figure~\ref{fig:mhcomp}, we show a comparison between the SALT derived total hydrogen mass outflow rates with those derived in \citet{xu2022}. We find a good agreement (accounting for the large SALT error bars) between the two methods, which is surprising  given the discrepancies found in the individual ionic column densities. We believe the agreement is fortuitous, as it is a consequence of the different methods used to compute the MORs. As we said in Section~\ref{sec:quantity_definitions}, we compute the MORs at $R_{SF}$, from Equation~\ref{eq:Mass}.  \citet{xu2022}, on the other hand, computes the MORs as follows: 
\begin{equation}
\dot M_{out}= \Omega \mu m_p R_0 \int \frac{dN_H}{dv} v dv,
\label{eq:xumor}
\end{equation}

\noindent
where $\Omega$ is assumed to be $4\pi$, $\frac{dN_H}{dv} v dv$ is the total hydrogen column density per observed velocity bin, and $R_0=2R_{SF}$.
The agreement between the two estimates of the MORs is likely due to the fact that the initial velocities that enter in Equation~\ref{eq:Mass}, are much smaller than the typical outflow velocities in \citet{xu2022}, on occasion even by an order of magnitude.
Additionally, SALT finds that the opening angle, $\alpha$, is not  $\pi/2$, resulting in smaller outflow rates than those obtained under the assumption of  $\Omega=4\pi$. 
Further discussion of the SALT mass outflow rates and their implications on galaxy properties will be included in a companion paper.

The difference in  velocities has implications on the energetic of the outflows: The lower SALT values suggest that the bulk of the outflow is located close to the galaxy, and moves relatively slowly. The measurement of the centroid velocities computed with a double Gaussian fitting, instead, assumes that  most of the gas in the outflow is located in a pseudo-shell, at a given distance and moving at one, much larger, specific velocity.

\section{Possible Origins of the Discrepancies}\label{sec:discussion}

In the following we perform a number of experiments to try and understand the origin of the observed discrepancy between the column densities derived using the canonical PCM method and the SALT modeling. 

\subsection{Comparison with Full Radiative Transfer Models}
To better understand the origin of the discrepancy, we compare the spectral predictions of both SALT and PCM to mock spectra generated with the numerical Monte Carlo based scattering code, RASCAS \citep{rascas2020}. The spectra were taken from \citet{carr2023}, contain the \sitwo 1190\AA, 1193\AA\ doublet, and were drawn from the same portion of the outflow parameter space as used in our model fitting. The base models assume simple bi-conical geometries and include line broadening due to turbulent motions of the gas.  In particular, we assume the Doppler parameter scales as $0.1 v(r)$ \citep{fielding2020}. Each spectrum was smoothed to a resolution of 20 km s$^{-1}$, or approximately the CLASSY value.  Gaussian noise was added until a signal-to-noise value of 10 (similar to the values measured in the CLASSY data) was achieved in the continuum.  

\begin{figure}[ht]
    \centering
    \includegraphics[width=0.99\linewidth]{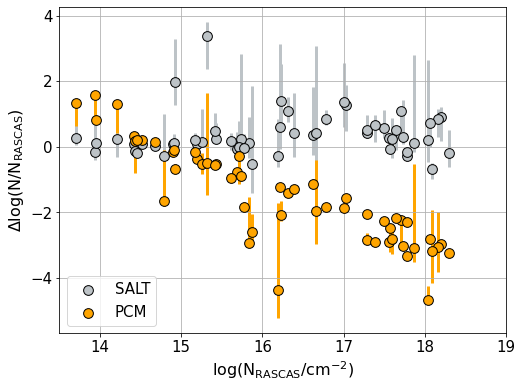}
    \caption{Results of the performance of the two methods (SALT and PCM) in recovering the true column densities for a number simulated data generated using RASCAS (see text for details). The negative correlation between $\log{N/N_{\rm RASCAS}}$ and $\log{N_{\rm RASCAS}}$ for the PCM estimates shows that the discrepancy between the recovered and true densities grows with the density, similar to the behavior observed in Figure~\ref{fig:3comp}.}
    \label{fig:rascascomp}
\end{figure}

Figure~\ref{fig:rascascomp} shows the comparison between the recovered column densities versus the known values from the simulations for both SALT and PCM. The SALT model shows a bias towards overestimating the column density (by about a factor of 3 on average) for Log(N$_{\rm RASCAS}$)$\gtrsim 16.5$.  Carr et al. (2023) showed that this  overestimate results from the fact that SALT does not account for thermal/turbulent line broadening due to the assumed validity of the Sobolev approximation. On the other hand, the PCM modeling tends to overestimate the column density at low Log(N$_{\rm RASCAS}$) and underestimate it at high Log(N$_{\rm RASCAS}$) even more than a factor of 100. Furthermore, the bias grows worse with increasing optical depth. These results are in agreement with \cite{delaCruz2021} who tested similar empirical methods on synthetic spectra derived from simulations of multi-phase outflows. Jennings et al (in prep), also found a large underestimate in the PCM at large column densities.

An important note: These simulations do not indicate that either method is `correct' or `incorrect', rather that both methods have separate biases that result in discrepancies that grow larger with column density.
The analysis performed on the RASCAS simulations suggests the discrepancy between the SALT and PCM CLASSY analyses is due to an underestimation of the PCM, but does not explain why the discrepancy arises. We argue that this discrepancy is due to a combination of 1) the inability of PCM to properly account for finite spectral resolution, and 2) the fact that the absorption profile is generated in front of an extended source. 

\subsection{The Effect of Spectral Resolution}
Observational systematic effects can affect the column density estimates. For example, in their seminal work, \cite{savage1991} already pointed out the bias introduced by the finite spectral resolution. To quantify this bias, we perform an experiment aimed at testing only this effect by simulating the absorption component\footnote{We do not consider re-emission in this experiment.} resulting from a point source behind a homologous spherical flow for various line of sight column densities.  Note that $f_c = 1$ in this experiment.  The fiducial outflow has the following parameters:  $\gamma = 1$, $\delta = 3$, $v_0 = 25$ \kmps, $v_{\infty} = 500 $\kmps. We do not add noise to these simulated spectra, which otherwise can be computed analytically (see Appendix~\ref{sec:salt_point_source} for details). 

\begin{figure}[ht]
    \centering 
    \includegraphics[width=1\linewidth]{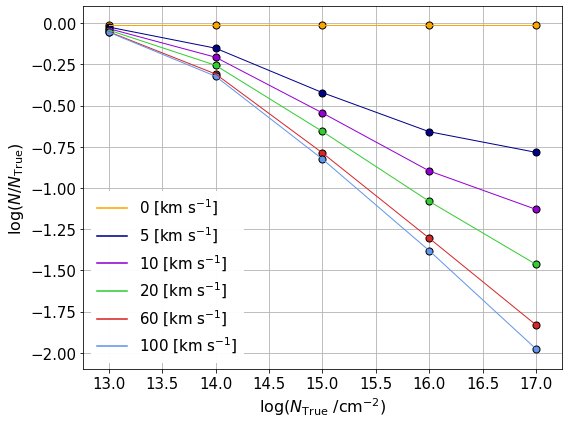}
    \caption{Estimates of the column density with the PCM (i.e., technique of direct integration) for a point source behind a homologous spherical flow according to column density and resolution.  The PCM fails to account for instrumental smoothing, and therefore cannot recover the correct column density.  The disagreement grows worse with increasing column density and resolution.  }
    \label{fig:point_source}
\end{figure}

To quantify the bias, we measured the ratio of the true column density, $N_{True}$, with the value recovered by the direct integration of the apparent optical depth, $N$.  We then repeated the experiment for different spectral resolutions by convolving the un-smoothed spectrum with a Gaussian kernel. The results are shown in Figure~\ref{fig:point_source}.  While the direct integration can perfectly recover the column density from the fully resolved profiles, it tends to diverge for the convolved ones.  Furthermore, the bias  grows worse with increasing column density, approaching nearly 1.5 orders of magnitude when $\log{(N \ [\rm cm^{-2}])} = 16-17$ near a resolution of 20 km/s or the resolution of CLASSY.  

The discrepancy occurs because the column density in the Apparent Optical Depth method is obtained by integrating over the optical depth, or $\tau = \ln{\left(I_0/I\right)}$.  While the EW of the un-smoothed line profile is conserved after convolution, the integration over $\tau$, generally, will not be, unless $\tau \ll 1$, so that $e^{-\tau} \rightarrow 1- \tau$  is a valid approximation. In this case, $\tau(v)$ adds linearly over the line profile, and N is directly proportional to the equivalent width of the line. The key point is that the effect is due to the combination of both optically thick absorption and low spectral resolution, explaining why the discrepancy between the two techniques is minimized at low column densities.  

\subsection{Absorption Against Extended Sources}
An additional effect to consider is the extended nature of the CLASSY galaxies. This extent implies that the overall optical depth derived from the fitting of a line profile is an average over different lines of sight. This average can introduce a bias because,  by Jensens' inequality, 

\begin{eqnarray}
    e^{-\tau_{\rm avg}}\leq \frac{e^{-\tau_1}+ ...+ e^{-\tau_n}}{n},
    \label{Jensen's_Inequality}
\end{eqnarray}
where the summation is over different lines of sight.  If the inequality is strict, then the PCM model may predict a smaller value of $\tau$ to account for the weaker absorption that occurs over multiple lines of sight. For the idealized models considered in this work, each line of sight will have equal weighting; however, this may not be the case in reality where the luminosity is not uniform across the emitting region.  Understanding the impact this type of error has on the interpretation of real data is better suited for tests against resolved simulations of galaxies (e.g., Jennings et al., in prep), and we do not attempt to quantify it here.

\section{Summary and Conclusions}\label{conclusion}
The UV spectra of galaxies include a wealth of absorption lines of  low and high ionization metals that can be used to infer the properties of the outflows which are almost ubiquitous around starburst galaxies. These lines are canonically modeled using variations of the empirical Apparent Optical Depth method, which enables the modeling of a velocity-dependent covering fraction. This approach neglects the spatial density profile of the outflowing gas, possibly resulting in an underestimate of the column density if a substantial amount of dense material is concentrated in a small velocity range. Semi-analytical models on the other hand relax this assumption, assuming that the density and velocity profiles are power law. Comparing the two approaches enable an assessment of the systematic uncertainties introduced by the modeling technique.

Here we compared the results of PCM and SALT modeling applied to the UV spectra of outflows identified in 17 CLASSY galaxies. We use the PCM results of \citet{xu2022}. For SALT, we model the FUV resonant transitions of three ionization states of silicon and constrain the outflow ionic column densities, geometry, and mass outflow rates. 
We find that the column densities estimated with the two methods do not agree with each other, and that the discrepancy increases with the SALT column density, up to two orders of magnitude for the most optically thick outflows.  

Using simulated spectra created with the RASCAS radiative transfer code as reference, we conclude that the reason for this discrepancy is that PCM misses the contribution of dense gas at low velocities, due to the finite resolution of the spectra.  SALT is less affected by finite-resolution effects because it computes the  outflow model which is then smoothed to the known instrumental resolution. The comparison with RASCAS shows that SALT's column densities agree very well with the simulated ones, although when the true column density is larger than $\approx log(N_{true}=16)$ they can be over-predicted by up to a factor of three because SALT neglects the outflow's thermal and turbulent velocities. While the SALT model takes into account the sources' extents, it is not immediately clear how well it can account for variations in the optical depth along different lines of sight.  

The results of this paper highlight a possible bias in the canonical methods of deriving outflow column densities, possibly resulting from the existence of a substantial amount of dense gas at low velocities. Whether this gas is actually present in galactic outflows remain to be confirmed. Numerical simulations do predict that outflow density and velocity profiles are not constant with radius \citep[e.g.,][]{muratov2015,schneider2020}, although not necessarily as power laws as assumed in the SALT formalism. This important issue can be addressed with the combination of the Habitable Worlds Telescope and a high-resolution, UV, Integral Field Unit spectrograph that would provide the emission and absorption line maps that are crucial to constrain this problem.

\section{Acknowledgement}
C.C. would like to acknowledge Harley Katz and Chris Howk for insightful discussions. Support for this research was provided by the University of Minnesota Undergraduate Research Opportunities Program Grant (UROP). We acknowledge the Minnesota Supercomputing Institute (MSI) at the University of Minnesota for providing the computational resources used in this project. Based on observations with the NASA/ESA Hubble Space Telescope obtained at the Space Telescope Science Institute, which is operated by the Association of Universities for Research in Astronomy, Incorporated, under NASA contract NAS5-26555. Support for Program number HST-AR-15792 was provided through a grant from the STScI under NASA contract NAS5-26555.

\clearpage

\appendix

\section{Example Corner Plot}\label{app:corner}
\begin{figure*}[!htb]
    \centering
    \includegraphics[width=.95\linewidth]{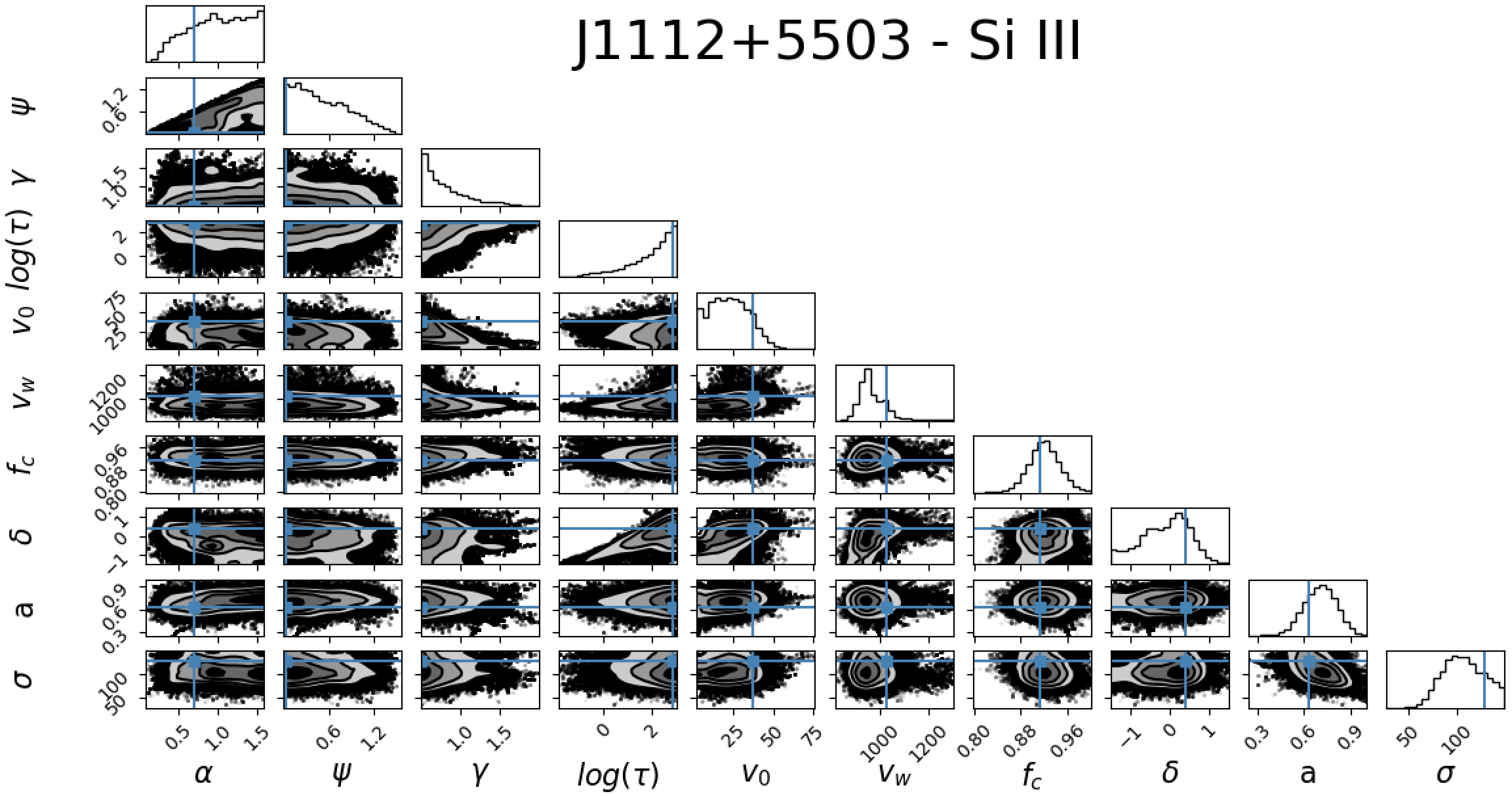}
    \caption{Corner plot for J1112+5503 for the \sithree\ line fit. Blue points represent the chosen maximum likelihood values.}
    \label{fig:corner}
\end{figure*}

The example corner plot in Figure \ref{fig:corner} is included as an example of the sampled parameter space using \textit{emcee}. Note that the reason the an individual best fit parameter does not necessarily line up with the peak of the marginalized distribution is that the position of the maximum likelihood is determined in the N-parameter space, not in the individual N marginalized distributions.

\clearpage

\section{SALT Model Fit Parameters}\label{app:fits}
Table~\ref{tab:pars1} displays all SALT fit parameters. 

\begin{table*}[h]

 \caption{SALT parameters for CLASSY galaxies. $\alpha$, $\psi$, $\gamma$, $\delta$, $\tau_0$, $f_c$, $v_0$, and $v_w$ are free parameters in SALT. $v_{ap}$ is derived from Table~\ref{tab:pars}. $k$ is derived from the dust optical depth (see Table~\ref{tab:pars}). $n_0$, $R_W$, N, and $\dot M$, are derived from Equations~\ref{eq:n0}, \ref{eq:rw}, \ref{eq:NN} and \ref{eq:Mass} respectively. The computation of $R_{COS}$ is described in Section~\ref{sec:data}. $R_{SF}$ is taken from \citet{xu2022}. Upper and lower uncertainties are described in Section~\ref{sec:quantity_definitions}. Galaxies denoted with a `$\dagger$' symbol did not pass the `F-test' in \citet{xu2022} indicating that an outflow model was even an improvement over a purely static ISM model (i.e. that the galaxy even exhibited outflow behavior.) The specific ionization line(s) that failed to pass are also indicated.}
 \label{tab:pars1}

 \centering
 \scalebox{0.7}{
 \begin{tabular}{*{22}{c}}
 
  \hline
  Name & Ion & $\alpha$ & $\psi$ & $\gamma$ & $\delta$ & log($\tau_0$) & $f_c$ & log(k) & $v_0$ & $v_w$ & $v_{ap}$& log($n_0$)& log(N) & log($\dot M$) & $R_W$ & $R_{COS}$ & $R_{SF}$\\
  &  & ($^{\circ}$) &($^{\circ}$) & && & &  & (km/s) &  (km/s)&  (km/s)& ($cm^{-3}$) & ($cm^{-2}$) & ($M_{\odot} yr^{-1}$) & (kpc) & (kpc) & (kpc)\\
  \hline
J0021+0052 & Si II & $38^{+8}_{-1} $ & $ 51^{+4}_{-2} $ & $ 1.9^{+0.1}_{-0.2} $ & $ 3.1^{+1.0}_{-0.5} $ & $ -0.1^{+1.0}_{-0.3} $ & $ 0.96^{+0.02}_{-0.32} $ & $ -0.5^{+0.2}_{-0.2} $ & $ 92^{+23}_{-12} $ & $ 915^{+163}_{-141} $ & $ 628^{+214}_{-181} $ & $ -4.1^{+1.9}_{-0.7} $ & $15.7^{+1.3}_{-0.5} $ & $-2.6^{+1.7}_{-0.7} $ &$1.5^{+0.4}_{-0.2}$\\
 & Si III & $40^{+16}_{-8} $ & $ 12^{+12}_{-7} $ & $ 1.0^{+0.2}_{-0.0} $ & $ 1.8^{+0.7}_{-0.1} $ & $ 0.8^{+0.9}_{-0.4} $ & $ 0.7^{+0.03}_{-0.04} $ & $ -0.8^{+0.2}_{-0.2} $ & $ 2^{+1}_{-0} $ & $ 813^{+5}_{-45} $ & $ 12^{+13}_{-4} $ & $ -4.7^{+1.0}_{-0.8} $ & $16.1^{+0.9}_{-0.6} $ & $-4.6^{+1.0}_{-0.8} $ &$33.5^{+45.4}_{-22.3}$&1.27&0.45\\
 & Si IV & $88^{+0}_{-27} $ & $ 1^{+20}_{-0} $ & $ 0.6^{+0.4}_{-0.0} $ & $ 1.1^{+0.8}_{-0.0} $ & $ -0.8^{+1.6}_{-0.3} $ & $ 0.63^{+0.03}_{-0.04} $ & $ -1.1^{+0.3}_{-0.4} $ & $ 5^{+5}_{-2} $ & $ 541^{+4}_{-6} $ & $ 24^{+19}_{-13} $ & $ -5.2^{+1.1}_{-1.0} $ & $15.8^{+1.0}_{-0.7} $ & $-4.6^{+1.1}_{-0.9} $ &$43.6^{+274.3}_{-35.2}$\\
\hline
J0036-3333 & Si II & $47^{+3}_{-11} $ & $ 19^{+5}_{-10} $ & $ 1.1^{+0.3}_{-0.2} $ & $ 1.9^{+0.8}_{-0.2} $ & $ -1.4^{+1.0}_{-0.2} $ & $ 0.57^{+0.03}_{-0.09} $ & $ -0.7^{+0.2}_{-0.2} $ & $ 60^{+6}_{-19} $ & $ 576^{+139}_{-21} $ & $ 133^{+60}_{-47} $ & $ -4.9^{+1.1}_{-0.6} $ & $14.9^{+0.7}_{-0.2} $ & $-5.1^{+0.7}_{-0.4} $ &$1.2^{+1.7}_{-0.5}$\\
 & Si III & $59^{+13}_{-16} $ & $ 35^{+12}_{-22} $ & $ 0.6^{+0.4}_{-0.0} $ & $ 1.7^{+0.6}_{-0.1} $ & $ 1.4^{+0.6}_{-0.4} $ & $ 0.96^{+0.01}_{-0.01} $ & $ -0.9^{+0.2}_{-0.2} $ & $ 2^{+0}_{-0} $ & $ 584^{+53}_{-18} $ & $ 7^{+5}_{-2} $ & $ -3.9^{+0.9}_{-0.6} $ & $16.5^{+0.6}_{-0.5} $ & $-4.7^{+1.0}_{-0.6} $ &$54.7^{+754.0}_{-50.0}$&0.29&0.12\\
 & Si IV & $61^{+3}_{-7} $ & $ 7^{+11}_{-5} $ & $ 2.0^{+0.0}_{-0.2} $ & $ 4.0^{+0.3}_{-0.5} $ & $ 0.7^{+0.7}_{-0.2} $ & $ 0.97^{+0.01}_{-0.08} $ & $ -0.6^{+0.1}_{-0.1} $ & $ 15^{+3}_{-5} $ & $ 1233^{+0}_{-17} $ & $ 68^{+23}_{-39} $ & $ -4.0^{+1.0}_{-0.4} $ & $16.1^{+0.9}_{-0.2} $ & $-4.1^{+0.5}_{-0.3} $ &$1.6^{+1.4}_{-0.4}$\\
\hline
J0405-3648$^\dagger$ & Si II$^\dagger$ & $44^{+21}_{-21} $ & $ 36^{+29}_{-18} $ & $ 0.6^{+0.7}_{-0.0} $ & $ 4.0^{+0.5}_{-1.1} $ & $ -1.1^{+1.3}_{-0.5} $ & $ 0.8^{+0.09}_{-0.34} $ & $ -0.8^{+0.2}_{-0.2} $ & $ 37^{+54}_{-21} $ & $ 1037^{+128}_{-527} $ & $ 2^{+7}_{-1} $ & $ -5.6^{+1.9}_{-1.1} $ & $14.1^{+1.3}_{-0.8} $ & $-5.1^{+1.8}_{-1.1} $ &$3.7^{+18.7}_{-2.3}$\\
 & Si III$^\dagger$ & $51^{+20}_{-13} $ & $ 36^{+18}_{-20} $ & $ 1.9^{+0.0}_{-0.7} $ & $ 3.2^{+0.7}_{-0.6} $ & $ -1.9^{+1.3}_{-0.1} $ & $ 0.8^{+0.08}_{-0.22} $ & $ -0.8^{+0.1}_{-0.2} $ & $ 33^{+11}_{-13} $ & $ 973^{+161}_{-432} $ & $ 1^{+4}_{-1} $ & $ -5.8^{+2.3}_{-0.8} $ & $14.5^{+1.4}_{-0.6} $ & $-4.8^{+1.9}_{-0.8} $ &$4.1^{+19.6}_{-2.5}$&0.04&0.37\\
 & Si IV$^\dagger$ & $61^{+16}_{-26} $ & $ 44^{+11}_{-25} $ & $ 1.4^{+0.3}_{-0.4} $ & $ 3.3^{+0.8}_{-0.8} $ & $ -0.5^{+1.4}_{-0.8} $ & $ 0.43^{+0.18}_{-0.07} $ & $ -0.8^{+0.1}_{-0.3} $ & $ 26^{+15}_{-7} $ & $ 746^{+267}_{-339} $ & $ 1^{+4}_{-1} $ & $ -5.5^{+2.5}_{-0.9} $ & $14.7^{+2.1}_{-0.7} $ & $-4.6^{+2.0}_{-0.9} $ &$3.3^{+13.3}_{-1.7}$\\
\hline
J0926+4427 & Si II & $76^{+1}_{-3} $ & $ 72^{+1}_{-2} $ & $ 2.0^{+0.0}_{-0.1} $ & $ 2.8^{+1.0}_{-0.3} $ & $ 0.3^{+0.6}_{-0.2} $ & $ 0.43^{+0.05}_{-0.04} $ & $ -0.3^{+0.1}_{-0.1} $ & $ 30^{+7}_{-3} $ & $ 444^{+11}_{-8} $ & $ 303^{+72}_{-56} $ & $ -4.4^{+0.6}_{-0.5} $ & $16.1^{+0.6}_{-0.4} $ & $-2.7^{+0.6}_{-0.5} $ &$2.6^{+0.4}_{-0.3}$\\
 & Si III & $83^{+2}_{-19} $ & $ 73^{+3}_{-47} $ & $ 0.9^{+0.2}_{-0.2} $ & $ 4.3^{+0.2}_{-0.5} $ & $ 0.4^{+0.4}_{-1.0} $ & $ 0.79^{+0.08}_{-0.04} $ & $ -0.2^{+0.1}_{-0.1} $ & $ 70^{+43}_{-13} $ & $ 897^{+162}_{-138} $ & $ 293^{+106}_{-79} $ & $ -5.2^{+0.6}_{-0.4} $ & $15.5^{+0.5}_{-0.4} $ & $-2.9^{+0.5}_{-0.4} $ &$7.5^{+10.8}_{-3.4}$&2.13&0.66\\
 & Si IV & $79^{+3}_{-26} $ & $ 70^{+3}_{-40} $ & $ 0.8^{+0.3}_{-0.1} $ & $ 4.1^{+0.3}_{-0.6} $ & $ 0.9^{+0.5}_{-0.8} $ & $ 0.8^{+0.07}_{-0.07} $ & $ -0.3^{+0.1}_{-0.1} $ & $ 86^{+21}_{-20} $ & $ 1295^{+2}_{-328} $ & $ 250^{+112}_{-74} $ & $ -4.4^{+0.7}_{-0.6} $ & $16.1^{+0.7}_{-0.5} $ & $-2.5^{+0.6}_{-0.5} $ &$9.9^{+14.9}_{-5.3}$\\
\hline
J0934+5514$^\dagger$ & Si II & $20^{+17}_{-10} $ & $ 68^{+11}_{-17} $ & $ 1.2^{+0.4}_{-0.4} $ & $ 3.5^{+0.7}_{-0.8} $ & $ -1.7^{+2.2}_{-0.1} $ & $ 0.87^{+0.06}_{-0.52} $ & $ 0.4^{+0.2}_{-0.2} $ & $ 60^{+45}_{-33} $ & $ 972^{+161}_{-373} $ & $ 24^{+21}_{-18} $ & $ -3.7^{+1.7}_{-1.7} $ & $13.8^{+2.0}_{-1.4} $ & $-5.1^{+1.9}_{-2.0} $ &$0.5^{+2.9}_{-0.3}$\\
 & Si III$^\dagger$ & $2^{+34}_{-1} $ & $ 0^{+53}_{-0} $ & $ 1.0^{+0.5}_{-0.3} $ & $ 2.0^{+1.6}_{-0.2} $ & $ -1.4^{+2.1}_{-0.3} $ & $ 0.14^{+0.44}_{-0.06} $ & $ 0.4^{+0.1}_{-0.2} $ & $ 120^{+13}_{-60} $ & $ 970^{+154}_{-440} $ & $ 27^{+24}_{-17} $ & $ -3.7^{+1.8}_{-1.8} $ & $14.4^{+2.4}_{-1.4} $ & $-4.6^{+2.2}_{-2.0} $ &$0.5^{+1.8}_{-0.3}$&0.04&0.08\\
 & Si IV & $82^{+3}_{-36} $ & $ 87^{+0}_{-51} $ & $ 1.4^{+0.3}_{-0.4} $ & $ 3.6^{+0.7}_{-1.0} $ & $ -1.2^{+1.6}_{-0.4} $ & $ 0.51^{+0.28}_{-0.29} $ & $ 0.3^{+0.2}_{-0.3} $ & $ 57^{+5}_{-16} $ & $ 632^{+221}_{-306} $ & $ 14^{+8}_{-6} $ & $ -4.6^{+2.1}_{-1.1} $ & $14.7^{+1.9}_{-1.0} $ & $-5.2^{+2.0}_{-1.0} $ &$0.4^{+1.3}_{-0.2}$\\
\hline
J0938+5428 & Si II & $54^{+5}_{-1} $ & $ 63^{+3}_{-1} $ & $ 1.4^{+0.3}_{-0.2} $ & $ 3.8^{+0.6}_{-1.1} $ & $ 1.9^{+0.5}_{-0.9} $ & $ 0.91^{+0.04}_{-0.12} $ & $ 0.1^{+0.2}_{-0.2} $ & $ 102^{+13}_{-18} $ & $ 1001^{+111}_{-440} $ & $ 370^{+94}_{-63} $ & $ -3.0^{+0.9}_{-1.1} $ & $17.3^{+0.9}_{-1.1} $ & $-1.1^{+0.9}_{-1.1} $ &$2.5^{+0.9}_{-1.1}$\\
 & Si III & $31^{+7}_{-9} $ & $ 1^{+5}_{-0} $ & $ 1.3^{+0.3}_{-0.5} $ & $ 3.6^{+0.5}_{-0.5} $ & $ -2.0^{+1.1}_{-0.0} $ & $ 0.94^{+0.03}_{-0.25} $ & $ 0.1^{+0.1}_{-0.2} $ & $ 131^{+8}_{-47} $ & $ 547^{+60}_{-44} $ & $ 219^{+177}_{-127} $ & $ -5.6^{+1.9}_{-0.7} $ & $14.6^{+1.4}_{-0.4} $ & $-4.3^{+1.9}_{-0.6} $ &$4.5^{+16.7}_{-2.9}$&1.32&0.51\\
 & Si IV & $26^{+33}_{-5} $ & $ 4^{+29}_{-2} $ & $ 0.8^{+0.6}_{-0.1} $ & $ 2.0^{+1.8}_{-0.2} $ & $ -1.8^{+1.5}_{-0.1} $ & $ 0.53^{+0.16}_{-0.13} $ & $ 0.1^{+0.1}_{-0.2} $ & $ 106^{+17}_{-48} $ & $ 1277^{+11}_{-223} $ & $ 210^{+170}_{-107} $ & $ -5.1^{+0.9}_{-0.7} $ & $14.9^{+0.8}_{-0.4} $ & $-3.8^{+0.9}_{-0.7} $ &$5.2^{+12.7}_{-2.6}$\\
\hline
J0940+2935 & Si II & $40^{+26}_{-15} $ & $ 26^{+23}_{-15} $ & $ 0.6^{+0.7}_{-0.1} $ & $ 3.7^{+0.6}_{-0.9} $ & $ 1.6^{+0.8}_{-2.3} $ & $ 0.97^{+0.01}_{-0.37} $ & $ -0.3^{+0.1}_{-0.3} $ & $ 8^{+20}_{-2} $ & $ 1029^{+125}_{-452} $ & $ 3^{+7}_{-2} $ & $ -5.0^{+2.3}_{-1.0} $ & $14.8^{+1.6}_{-1.0} $ & $-5.4^{+2.1}_{-0.9} $ &$1.7^{+13.6}_{-1.1}$\\
 & Si III & $72^{+8}_{-21} $ & $ 34^{+19}_{-17} $ & $ 0.5^{+0.7}_{-0.0} $ & $ 3.0^{+1.0}_{-0.4} $ & $ 2.9^{+0.1}_{-3.0} $ & $ 0.99^{+0.01}_{-0.27} $ & $ -0.3^{+0.1}_{-0.2} $ & $ 3^{+12}_{-0} $ & $ 1028^{+138}_{-494} $ & $ 2^{+6}_{-1} $ & $ -5.1^{+2.0}_{-1.1} $ & $14.8^{+1.6}_{-0.9} $ & $-5.4^{+1.9}_{-1.1} $ &$2.5^{+28.8}_{-1.8}$&0.02&0.12\\
 & Si IV & $68^{+11}_{-18} $ & $ 55^{+10}_{-31} $ & $ 2.0^{+0.0}_{-0.6} $ & $ 2.8^{+0.9}_{-0.4} $ & $ -0.6^{+1.1}_{-0.8} $ & $ 0.43^{+0.18}_{-0.08} $ & $ -0.4^{+0.2}_{-0.2} $ & $ 13^{+17}_{-4} $ & $ 1243^{+28}_{-568} $ & $ 2^{+5}_{-1} $ & $ -5.2^{+1.5}_{-0.8} $ & $14.5^{+1.5}_{-0.4} $ & $-5.4^{+1.2}_{-0.6} $ &$1.3^{+3.6}_{-0.7}$\\
\hline
J1024+0524 & Si II & $77^{+1}_{-1} $ & $ 87^{+1}_{-1} $ & $ 0.9^{+0.2}_{-0.1} $ & $ 1.7^{+1.1}_{-0.2} $ & $ -1.1^{+2.0}_{-0.2} $ & $ 0.93^{+0.03}_{-0.07} $ & $ -0.1^{+0.2}_{-0.3} $ & $ 82^{+6}_{-10} $ & $ 785^{+262}_{-383} $ & $ 134^{+14}_{-13} $ & $ -3.7^{+1.5}_{-1.5} $ & $17.1^{+1.3}_{-1.7} $ & $-2.2^{+1.4}_{-1.4} $ &$2.5^{+3.4}_{-1.9}$\\
 & Si III & $41^{+25}_{-6} $ & $ 10^{+20}_{-5} $ & $ 1.8^{+0.1}_{-0.7} $ & $ 3.7^{+0.6}_{-0.9} $ & $ -0.5^{+1.2}_{-0.5} $ & $ 0.94^{+0.03}_{-0.09} $ & $ 0.0^{+0.2}_{-0.2} $ & $ 18^{+6}_{-8} $ & $ 317^{+30}_{-17} $ & $ 29^{+19}_{-16} $ & $ -4.9^{+1.1}_{-1.0} $ & $15.6^{+1.0}_{-0.7} $ & $-4.3^{+1.0}_{-0.9} $ &$4.2^{+11.4}_{-2.5}$&0.46&0.27\\
 & Si IV & $46^{+22}_{-10} $ & $ 21^{+18}_{-11} $ & $ 2.0^{+0.0}_{-0.6} $ & $ 4.6^{+0.3}_{-1.2} $ & $ 1.1^{+0.9}_{-0.8} $ & $ 0.77^{+0.07}_{-0.06} $ & $ 0.0^{+0.1}_{-0.2} $ & $ 9^{+6}_{-4} $ & $ 623^{+317}_{-162} $ & $ 20^{+19}_{-12} $ & $ -4.3^{+0.9}_{-0.9} $ & $15.9^{+0.9}_{-0.8} $ & $-4.0^{+0.7}_{-0.8} $ &$5.9^{+31.6}_{-3.4}$\\
\hline
J1025+3622 & Si II & $56^{+6}_{-4} $ & $ 4^{+7}_{-2} $ & $ 1.5^{+0.2}_{-0.2} $ & $ 2.4^{+0.7}_{-0.2} $ & $ -1.2^{+0.8}_{-0.1} $ & $ 0.71^{+0.04}_{-0.08} $ & $ -0.6^{+0.2}_{-0.1} $ & $ 58^{+6}_{-16} $ & $ 458^{+29}_{-20} $ & $ 131^{+42}_{-43} $ & $ -5.7^{+0.7}_{-0.4} $ & $15.2^{+0.5}_{-0.3} $ & $-3.9^{+0.6}_{-0.3} $ &$3.6^{+1.7}_{-0.8}$\\
 & Si III & $48^{+21}_{-10} $ & $ 14^{+18}_{-7} $ & $ 0.6^{+0.5}_{-0.1} $ & $ 3.4^{+0.7}_{-0.7} $ & $ 0.3^{+1.2}_{-0.7} $ & $ 0.72^{+0.08}_{-0.05} $ & $ -0.5^{+0.1}_{-0.2} $ & $ 64^{+19}_{-36} $ & $ 708^{+256}_{-155} $ & $ 74^{+69}_{-43} $ & $ -4.7^{+1.2}_{-1.0} $ & $16.1^{+1.3}_{-1.0} $ & $-2.9^{+1.1}_{-1.0} $ &$14.5^{+58.0}_{-9.1}$&1.58&0.79\\
 & Si IV & $81^{+4}_{-22} $ & $ 50^{+8}_{-32} $ & $ 1.9^{+0.1}_{-0.6} $ & $ 2.9^{+0.8}_{-0.6} $ & $ 0.3^{+1.1}_{-0.6} $ & $ 0.9^{+0.03}_{-0.09} $ & $ -0.6^{+0.2}_{-0.2} $ & $ 11^{+9}_{-3} $ & $ 512^{+260}_{-26} $ & $ 37^{+31}_{-17} $ & $ -4.8^{+1.0}_{-1.0} $ & $16.2^{+0.9}_{-0.8} $ & $-3.2^{+0.8}_{-0.9} $ &$17.0^{+79.3}_{-9.4}$\\
\hline
J1105+4444 & Si II & $35^{+28}_{-9} $ & $ 20^{+16}_{-11} $ & $ 1.2^{+0.2}_{-0.3} $ & $ 4.2^{+0.3}_{-0.8} $ & $ 2.0^{+0.4}_{-1.0} $ & $ 0.91^{+0.04}_{-0.04} $ & $ -0.5^{+0.1}_{-0.2} $ & $ 12^{+10}_{-2} $ & $ 333^{+103}_{-34} $ & $ 2^{+1}_{-1} $ & $ -4.6^{+0.9}_{-0.9} $ & $16.9^{+0.8}_{-1.2} $ & $-2.4^{+0.9}_{-0.8} $ &$33.6^{+140.1}_{-23.1}$\\
 & Si III & $54^{+18}_{-11} $ & $ 17^{+20}_{-8} $ & $ 1.9^{+0.0}_{-0.8} $ & $ 4.5^{+0.3}_{-1.4} $ & $ -0.8^{+2.1}_{-0.4} $ & $ 0.99^{+0.0}_{-0.07} $ & $ -0.6^{+0.2}_{-0.3} $ & $ 42^{+10}_{-20} $ & $ 270^{+68}_{-18} $ & $ 3^{+5}_{-2} $ & $ -4.8^{+1.2}_{-1.4} $ & $16.7^{+1.1}_{-1.4} $ & $-2.3^{+1.3}_{-1.3} $ &$15.0^{+68.0}_{-8.6}$&0.31&1.8\\
 & Si IV & $57^{+15}_{-15} $ & $ 33^{+16}_{-17} $ & $ 1.1^{+0.3}_{-0.3} $ & $ 4.3^{+0.3}_{-0.8} $ & $ -0.6^{+1.0}_{-0.3} $ & $ 0.84^{+0.04}_{-0.05} $ & $ -0.5^{+0.1}_{-0.2} $ & $ 48^{+4}_{-25} $ & $ 592^{+332}_{-159} $ & $ 3^{+4}_{-2} $ & $ -5.6^{+0.9}_{-0.6} $ & $15.5^{+0.8}_{-0.5} $ & $-3.2^{+0.7}_{-0.6} $ &$44.5^{+225.9}_{-28.7}$\\
\hline
J1112+5503 & Si II & $70^{+3}_{-6} $ & $ 18^{+8}_{-9} $ & $ 1.2^{+0.1}_{-0.1} $ & $ 4.4^{+0.1}_{-0.4} $ & $ 3.0^{+0.0}_{-0.2} $ & $ 0.7^{+0.01}_{-0.01} $ & $ 0.5^{+0.0}_{-0.1} $ & $ 30^{+2}_{-3} $ & $ 1165^{+33}_{-27} $ & $ 125^{+20}_{-18} $ & $ -2.4^{+0.2}_{-0.3} $ & $18.2^{+0.2}_{-0.3} $ & $-1.1^{+0.3}_{-0.4} $ &$10.9^{+5.8}_{-3.2}$\\
 & Si III & $40^{+25}_{-10} $ & $ 2^{+24}_{-1} $ & $ 0.5^{+0.2}_{-0.0} $ & $ 2.9^{+0.3}_{-0.5} $ & $ 2.8^{+0.1}_{-0.8} $ & $ 0.91^{+0.02}_{-0.01} $ & $ 0.3^{+0.1}_{-0.2} $ & $ 37^{+4}_{-17} $ & $ 1027^{+31}_{-81} $ & $ 57^{+25}_{-26} $ & $ -3.2^{+0.8}_{-1.1} $ & $17.5^{+0.8}_{-1.0} $ & $-2.2^{+1.0}_{-1.3} $ &$84.3^{+220.3}_{-59.4}$&1.64&0.47\\
 & Si IV & $71^{+5}_{-9} $ & $ 5^{+13}_{-2} $ & $ 1.1^{+0.2}_{-0.2} $ & $ 4.4^{+0.2}_{-0.5} $ & $ 0.8^{+0.4}_{-0.4} $ & $ 0.78^{+0.04}_{-0.06} $ & $ 0.5^{+0.1}_{-0.1} $ & $ 91^{+13}_{-18} $ & $ 1250^{+23}_{-28} $ & $ 347^{+145}_{-96} $ & $ -3.9^{+0.5}_{-0.5} $ & $16.6^{+0.4}_{-0.4} $ & $-2.1^{+0.4}_{-0.4} $ &$5.1^{+5.2}_{-1.9}$\\
\hline
J1150+1501 & Si II & $59^{+12}_{-36} $ & $ 33^{+33}_{-18} $ & $ 0.9^{+0.6}_{-0.2} $ & $ 4.3^{+0.4}_{-1.4} $ & $ -0.0^{+1.4}_{-1.1} $ & $ 0.98^{+0.01}_{-0.62} $ & $ 0.2^{+0.2}_{-0.3} $ & $ 12^{+55}_{-5} $ & $ 1297^{+1}_{-572} $ & $ 26^{+23}_{-17} $ & $ -4.1^{+2.1}_{-1.6} $ & $14.4^{+2.2}_{-1.7} $ & $-5.5^{+2.1}_{-1.4} $ &$0.5^{+2.2}_{-0.3}$\\
 & Si III & $46^{+25}_{-13} $ & $ 7^{+26}_{-3} $ & $ 0.5^{+0.6}_{-0.0} $ & $ 3.8^{+0.4}_{-0.6} $ & $ 0.9^{+0.9}_{-1.5} $ & $ 0.96^{+0.02}_{-0.15} $ & $ 0.3^{+0.1}_{-0.2} $ & $ 24^{+15}_{-11} $ & $ 1005^{+135}_{-577} $ & $ 10^{+12}_{-6} $ & $ -4.6^{+1.6}_{-1.0} $ & $15.1^{+1.3}_{-0.9} $ & $-5.2^{+1.4}_{-0.9} $ &$1.5^{+9.5}_{-1.1}$&0.04&0.07\\
 & Si IV & $50^{+21}_{-11} $ & $ 9^{+16}_{-5} $ & $ 1.3^{+0.4}_{-0.3} $ & $ 3.2^{+1.0}_{-0.7} $ & $ 1.8^{+0.4}_{-0.7} $ & $ 1.0^{+0.0}_{-0.04} $ & $ 0.2^{+0.1}_{-0.3} $ & $ 2^{+2}_{-0} $ & $ 962^{+172}_{-358} $ & $ 1^{+1}_{-0} $ & $ -3.3^{+0.6}_{-0.8} $ & $16.7^{+0.5}_{-0.6} $ & $-4.5^{+0.6}_{-0.7} $ &$3.8^{+46.5}_{-2.3}$\\
\hline
J1200+1343 & Si II & $62^{+2}_{-2} $ & $ 5^{+3}_{-2} $ & $ 2.0^{+0.0}_{-0.2} $ & $ 5.5^{+0.0}_{-0.6} $ & $ 2.5^{+0.1}_{-0.7} $ & $ 0.85^{+0.02}_{-0.01} $ & $ 0.1^{+0.0}_{-0.1} $ & $ 8^{+3}_{-0} $ & $ 281^{+26}_{-7} $ & $ 127^{+26}_{-22} $ & $ -3.4^{+0.3}_{-0.3} $ & $16.9^{+0.2}_{-0.3} $ & $-3.0^{+0.2}_{-0.4} $ &$1.4^{+0.2}_{-0.2}$\\
 & Si III & $29^{+5}_{-4} $ & $ 0^{+5}_{-0} $ & $ 1.0^{+0.2}_{-0.0} $ & $ 3.0^{+0.4}_{-0.2} $ & $ 2.9^{+0.0}_{-0.4} $ & $ 1.0^{+0.0}_{-0.02} $ & $ -0.2^{+0.1}_{-0.1} $ & $ 2^{+0}_{-0} $ & $ 632^{+93}_{-16} $ & $ 12^{+13}_{-3} $ & $ -3.3^{+0.3}_{-0.5} $ & $17.0^{+0.3}_{-0.4} $ & $-4.1^{+0.4}_{-0.5} $ &$28.8^{+25.3}_{-21.9}$&0.9&0.23\\
 & Si IV & $35^{+2}_{-2} $ & $ 1^{+2}_{-0} $ & $ 2.0^{+0.0}_{-0.1} $ & $ 4.4^{+0.3}_{-0.3} $ & $ 3.0^{+0.0}_{-0.3} $ & $ 0.99^{+0.01}_{-0.02} $ & $ -0.1^{+0.1}_{-0.1} $ & $ 2^{+1}_{-0} $ & $ 573^{+25}_{-27} $ & $ 39^{+25}_{-7} $ & $ -3.1^{+0.2}_{-0.4} $ & $17.1^{+0.2}_{-0.3} $ & $-3.7^{+0.4}_{-0.4} $ &$3.5^{+0.6}_{-0.7}$\\
\hline
J1314+3452 & Si II & $86^{+1}_{-9} $ & $ 65^{+4}_{-30} $ & $ 1.2^{+0.2}_{-0.2} $ & $ 4.7^{+0.2}_{-0.4} $ & $ 1.7^{+0.6}_{-0.4} $ & $ 0.95^{+0.02}_{-0.03} $ & $ -0.3^{+0.0}_{-0.1} $ & $ 10^{+4}_{-1} $ & $ 211^{+405}_{-11} $ & $ 32^{+15}_{-5} $ & $ -2.7^{+0.8}_{-0.4} $ & $16.6^{+0.3}_{-0.3} $ & $-4.1^{+0.5}_{-0.3} $ &$0.6^{+1.0}_{-0.5}$\\
 & Si III & $50^{+8}_{-8} $ & $ 0^{+11}_{-0} $ & $ 0.5^{+0.0}_{-0.0} $ & $ 1.8^{+0.3}_{-0.1} $ & $ 1.1^{+0.9}_{-0.5} $ & $ 1.0^{+0.0}_{-0.0} $ & $ -0.8^{+0.1}_{-0.2} $ & $ 2^{+0}_{-0} $ & $ 206^{+14}_{-3} $ & $ 3^{+0}_{-0} $ & $ -3.3^{+1.0}_{-1.0} $ & $16.5^{+0.8}_{-0.8} $ & $-5.9^{+1.0}_{-1.0} $ &$92.6^{+86.0}_{-57.8}$&0.04&0.02\\
 & Si IV & $54^{+11}_{-11} $ & $ 0^{+13}_{-0} $ & $ 0.5^{+0.1}_{-0.0} $ & $ 1.5^{+0.2}_{-0.1} $ & $ 0.1^{+0.5}_{-0.2} $ & $ 0.98^{+0.01}_{-0.03} $ & $ -1.1^{+0.1}_{-0.1} $ & $ 2^{+0}_{-0} $ & $ 216^{+35}_{-6} $ & $ 3^{+0}_{-0} $ & $ -4.5^{+0.7}_{-0.4} $ & $15.4^{+0.5}_{-0.3} $ & $-7.0^{+0.7}_{-0.5} $ &$95.7^{+138.5}_{-70.8}$\\
\hline
J1359+5726 & Si II & $39^{+35}_{-5} $ & $ 15^{+22}_{-7} $ & $ 0.9^{+0.4}_{-0.2} $ & $ 1.8^{+0.8}_{-0.2} $ & $ -0.5^{+1.0}_{-0.3} $ & $ 0.78^{+0.01}_{-0.01} $ & $ -0.7^{+0.2}_{-0.3} $ & $ 24^{+3}_{-5} $ & $ 399^{+11}_{-9} $ & $ 14^{+6}_{-5} $ & $ -5.2^{+1.6}_{-0.7} $ & $15.7^{+1.1}_{-0.3} $ & $-3.5^{+1.5}_{-0.7} $ &$12.9^{+42.7}_{-7.7}$\\
 & Si III & $87^{+1}_{-25} $ & $ 31^{+13}_{-18} $ & $ 0.8^{+0.5}_{-0.1} $ & $ 1.4^{+1.5}_{-0.1} $ & $ 0.0^{+2.3}_{-0.3} $ & $ 0.98^{+0.01}_{-0.02} $ & $ -0.6^{+0.2}_{-0.2} $ & $ 2^{+2}_{-0} $ & $ 460^{+53}_{-10} $ & $ 2^{+1}_{-1} $ & $ -3.9^{+0.7}_{-1.2} $ & $17.0^{+0.7}_{-0.7} $ & $-3.1^{+1.0}_{-1.2} $ &$63.9^{+600.9}_{-49.8}$&0.47&0.74\\
 & Si IV & $63^{+14}_{-12} $ & $ 55^{+9}_{-31} $ & $ 0.6^{+0.9}_{-0.1} $ & $ 1.8^{+1.1}_{-0.2} $ & $ 0.3^{+0.5}_{-0.4} $ & $ 0.82^{+0.03}_{-0.02} $ & $ -0.7^{+0.1}_{-0.3} $ & $ 8^{+3}_{-1} $ & $ 502^{+32}_{-33} $ & $ 5^{+3}_{-1} $ & $ -5.4^{+0.7}_{-0.5} $ & $15.8^{+0.7}_{-0.3} $ & $-4.0^{+0.6}_{-0.5} $ &$9.4^{+64.3}_{-3.7}$\\
\hline
J1428+1653 & Si II & $76^{+7}_{-3} $ & $ 73^{+4}_{-18} $ & $ 1.7^{+0.1}_{-0.2} $ & $ 4.2^{+0.5}_{-0.5} $ & $ 0.9^{+0.3}_{-0.5} $ & $ 0.95^{+0.03}_{-0.29} $ & $ -0.2^{+0.1}_{-0.1} $ & $ 19^{+8}_{-2} $ & $ 1115^{+86}_{-269} $ & $ 87^{+23}_{-19} $ & $ -4.9^{+0.5}_{-0.5} $ & $16.0^{+0.4}_{-0.3} $ & $-2.7^{+0.4}_{-0.5} $ &$8.7^{+3.4}_{-2.1}$\\
 & Si III & $51^{+20}_{-12} $ & $ 11^{+18}_{-5} $ & $ 1.0^{+0.3}_{-0.0} $ & $ 2.3^{+0.7}_{-0.2} $ & $ 1.5^{+0.6}_{-0.6} $ & $ 1.0^{+0.0}_{-0.02} $ & $ -0.4^{+0.1}_{-0.1} $ & $ 2^{+1}_{-0} $ & $ 818^{+3}_{-54} $ & $ 10^{+5}_{-3} $ & $ -4.8^{+0.8}_{-0.8} $ & $16.4^{+0.8}_{-0.6} $ & $-3.6^{+0.8}_{-0.8} $ &$59.3^{+65.8}_{-31.1}$&2.14&1.04\\
 & Si IV & $34^{+25}_{-6} $ & $ 7^{+16}_{-3} $ & $ 1.1^{+0.2}_{-0.0} $ & $ 2.7^{+0.3}_{-0.4} $ & $ 2.9^{+0.1}_{-0.7} $ & $ 0.96^{+0.02}_{-0.04} $ & $ -0.5^{+0.1}_{-0.2} $ & $ 2^{+0}_{-0} $ & $ 344^{+37}_{-27} $ & $ 7^{+3}_{-2} $ & $ -4.2^{+0.5}_{-0.8} $ & $17.1^{+0.4}_{-0.6} $ & $-3.2^{+0.6}_{-0.8} $ &$45.1^{+41.2}_{-23.1}$\\
\hline
J1429+0643 & Si II & $86^{+2}_{-4} $ & $ 58^{+6}_{-15} $ & $ 1.1^{+0.3}_{-0.1} $ & $ 3.5^{+0.6}_{-0.3} $ & $ 0.1^{+0.4}_{-0.3} $ & $ 0.64^{+0.05}_{-0.03} $ & $ -0.1^{+0.1}_{-0.1} $ & $ 55^{+9}_{-10} $ & $ 1033^{+121}_{-168} $ & $ 371^{+190}_{-94} $ & $ -4.7^{+0.4}_{-0.4} $ & $15.9^{+0.4}_{-0.3} $ & $-3.0^{+0.4}_{-0.4} $ &$4.5^{+2.6}_{-1.6}$\\
 & Si III & $55^{+3}_{-7} $ & $ 4^{+7}_{-2} $ & $ 0.6^{+0.4}_{-0.1} $ & $ 2.6^{+1.2}_{-0.3} $ & $ -1.6^{+1.0}_{-0.2} $ & $ 0.97^{+0.02}_{-0.06} $ & $ -0.1^{+0.1}_{-0.2} $ & $ 124^{+7}_{-37} $ & $ 1171^{+64}_{-231} $ & $ 396^{+322}_{-130} $ & $ -5.2^{+1.0}_{-0.6} $ & $15.4^{+0.8}_{-0.5} $ & $-3.5^{+0.7}_{-0.5} $ &$5.1^{+7.5}_{-2.8}$&2.06&0.44\\
 & Si IV & $54^{+4}_{-6} $ & $ 4^{+9}_{-2} $ & $ 0.8^{+0.3}_{-0.1} $ & $ 3.5^{+0.6}_{-0.4} $ & $ -0.7^{+0.7}_{-0.3} $ & $ 0.98^{+0.01}_{-0.1} $ & $ -0.2^{+0.1}_{-0.1} $ & $ 125^{+8}_{-28} $ & $ 606^{+176}_{-19} $ & $ 393^{+313}_{-120} $ & $ -4.8^{+0.7}_{-0.6} $ & $15.8^{+0.5}_{-0.4} $ & $-3.0^{+0.5}_{-0.5} $ &$4.0^{+4.8}_{-2.1}$\\

 \hline
 \end{tabular}%
 }

\end{table*}

\clearpage

\section{Effective Column Density}\label{app:column_density}

\begin{figure}[t]
	\centering
	\includegraphics[scale=0.48]{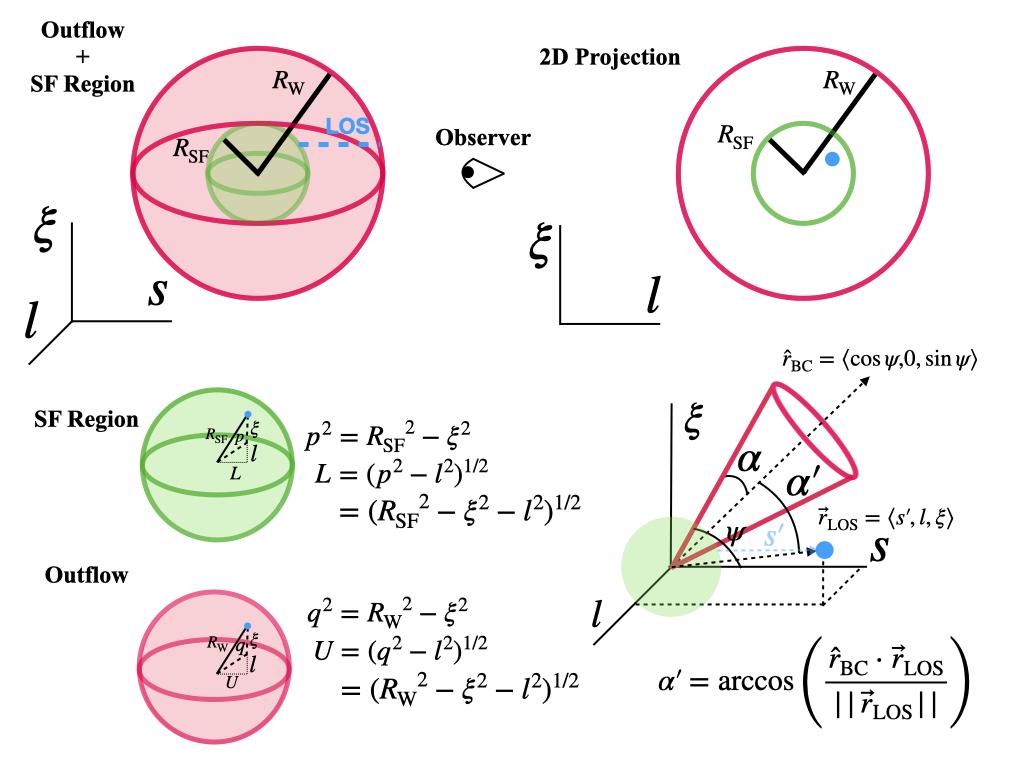}
	\caption{Schematic rendering illustrating how to compute the column density in the SALT formalism along an arbitrary line of sight.  \textbf{\emph{Top Left}}  A spherical outflow (salmon) surrounding a star forming region (light green).  We aim to compute the column density along the path shown in light blue.  \textbf{\emph{Top Right}} 2D projection of the outflow. \textbf{\emph{Bottom Left}}  To avoid integrating inside the source, we begin our calculation of the column density starting from the surface of the star forming region, or a distance $L$ from the $l\xi$-plane.  We compute the column density along this path until it terminates at the edge of the outflow a distance, $U$, from the $l\xi$-plane. \textbf{\emph{Bottom Right}}  We can account for a bi-conical outflow geometry by setting the density field to zero everywhere along the light blue path that falls outside the boundary of the bi-cone.  We can determine if a location along the path falls outside the bi-cone by checking if the angle subtended by the axis of the cone, $\hat{r}_{\rm BC}$, and the location of the point, $\vec{r}_{\rm LOS}$, denoted by $\alpha^{\prime}$, is greater than the opening angle of the cone, $\alpha$. }
	\label{fig:Neff}
\end{figure} 

Here we demonstrate how to compute the column density along an arbitrary sight line in the SALT formalism.  A diagram showcasing the details of the calculation is provided in Figure~\ref{fig:Neff}.  We begin with the case of a spherical outflow (i.e., $\alpha = 90^{\circ}$).  Consider a sight line starting on the surface of the star forming region and ending on the edge of the outflow.  See the light blue path in the upper left corner of Figure~\ref{fig:Neff}.  Let $(s^{\prime},\xi,l)$ be a point on this path defined by the coordinate system in Figure~\ref{fig:Neff}. It follows that the column density, $N_{(l,\xi),\rm sphere}$, along this path can be computed as  
\begin{eqnarray}
N_{(l,\xi),\rm sphere} = \int_L^U n_0\left(\frac{{s^{\prime}}^2+l^2+\xi^2}{R_\text{SF}^2}\right)^{-\delta/2}\text{d}s^{\prime} \, ,
\label{eq:N}
\end{eqnarray}
where the upper bound of integration is 
\begin{eqnarray}
U = (R_W^2-l^2-\xi^2)^{1/2},
\end{eqnarray}
and the lower bound of integration is
\begin{eqnarray}
L = (R_\text{SF}^2-l^2-\xi^2)^{1/2}.
\end{eqnarray}
Graphics detailing the calculations of $U$ and $L$ are provided in the lower left corner of Figure~\ref{fig:Neff}.

To generalize the spherical solution to that of a bi-cone, we force a bi-conical boundary condition by collapsing the integral in Equation~(\ref{eq:N}) to zero everywhere that the point $(s^{\prime},\xi,l)$ falls outside the bi-cone - that is, we construct an indicator function, $\Phi_{\text{bi-cone}}$, such that 
\begin{eqnarray}
N_{(l,\xi),\text{bi-cone}} = \int_{L}^{U}\Phi_{\text{bi-cone}}n_0\left(\frac{{s^{\prime}}^2+l^2+\xi^2}{R_\text{SF}^2}\right)^{-\delta/2}\text{d}s^{\prime},
\label{eq:N2}
\end{eqnarray}
where
\begin{gather} 
\Phi_{\text{bi-cone}} \equiv 
\begin{cases}
1 &\ \text{if} \  (s^{\prime},\xi,l) \in \text{bi-cone} \\[1em]
0 & \ \rm{otherwise}. \\[1em]
\end{cases}
\end{gather}
To determine if $(s^{\prime},\xi,l)$ is in the bi-cone, we first check if it lies in the top cone.  To do this, we check if the angle, $\alpha^{\prime}$, subtended by the axis of the cone and the ray, $\vec{r}_{\rm LOS} = \langle s^{\prime},\xi,l\rangle$, is greater than the opening angle of the bi-cone, $\alpha$.  See the bottom right corner of Figure~\ref{fig:Neff}.  Denoting the axis of the cone by the unit vector $\hat{r}_{\rm BC} = \langle \cos{\psi},0,\sin{\psi}\rangle$, we can compute $\alpha^{\prime}$ from the dot product as 
\begin{eqnarray}
    \alpha^{\prime} = \arccos{\left(\frac{ \hat{r}_{\rm BC} \cdot \vec{r}_{\rm LOS}}{||\vec{r}_{\rm LOS}||}\right)},
\end{eqnarray}
where $||\vec{r}_{\rm LOS}||$ is the magnitude of $\vec{r}_{\rm LOS}$.  If $\alpha^{\prime}>\alpha$, then $(s,l,\xi)$ is not in the top cone.  To check the bottom cone, we repeat the construction, but use $-\hat{r}_{BC}$.

\section{SALT Model with a Point Source}
\label{sec:salt_point_source}
To generate the spectra in Figure~\ref{fig:point_source}, we use a version of a spherical SALT model adapted to accommodate a point source.  The physical description of the outflow (i.e., velocity field, density field, etc.) remains unchanged between this version of SALT and former versions (e.g., \citealt{carr2018}).  However, because radiation will only be observed along a single line of sight, the equations used to compute the line profile will need to change.  For the work in this paper, we only consider the absorption profile.  Furthermore, for simplicity, we place the point source at the radial center of the outflow.  With these assumptions, we compute the normalized line profile for a point source behind an expanding spherical outflow as 
\begin{eqnarray}
    \frac{I}{I_0} = e^{-\tau_S},
\end{eqnarray}
where 
\begin{eqnarray}
    \tau_S = \frac{\tau_0}{\gamma}y^{(1-\gamma-\delta)/\gamma},
\end{eqnarray}
is the Sobolev optical depth defined along a single line of sight (see \citealt{carr2023} for a derivation) and
\begin{eqnarray}
    \tau_0 = \frac{\pi e^2}{m_e c}f_{lu}\lambda_{lu}n_0\frac{R_{SF}}{v_0}.
\end{eqnarray}
Note that the model assumes the outflow starts at a distance, $R_{SF}$, away from the point source.  In other words, the model is exactly the same as the spherical SALT model presented in \cite{carr2023}, but the spherical source has been replaced by a point source located at its center.

\clearpage
\newpage

\clearpage

\bibliography{refr}{}
\bibliographystyle{aasjournal}

\end{document}